\documentclass{emulateapj}

\usepackage{natbib}
\usepackage{amsmath}

\newcommand{\ugrizyjh}{\protect\hbox{$ugrizY\!\!J\!H$} }

\newcommand{\ubv}{\protect\hbox{$U\!BV$} }
\newcommand{\bvri}{\protect\hbox{$BV\!RI$} }
\newcommand{\bvrijh}{\protect\hbox{$BV\!RI\!J\!H$} }
\newcommand{\ubvrijhk}{\protect\hbox{$U\!BV\!RI\!J\!H\!K$} }
\newcommand{\bvrijhk}{\protect\hbox{$BV\!RI\!J\!H\!K$} }
\newcommand{\vrijhk}{\protect\hbox{$V\!RI\!J\!H\!K$} }
\newcommand{\vri}{\protect\hbox{$V\!RI$} }

\newcommand{\jh}{\protect\hbox{$J\!H$} }
\newcommand{\jhk}{\protect\hbox{$J\!H\!K$} }

\newcommand{\about}{$\sim\!\!$~}

\newcommand{\be}{\begin{displaymath}}
\newcommand{\ee}{\end{displaymath}}

\def\lsim{\hbox{\rlap{\raise 0.425ex\hbox{$<$}}\lower 0.65ex\hbox{$\sim$}}}
\def\gsim{\hbox{\rlap{\raise 0.425ex\hbox{$>$}}\lower 0.65ex\hbox{$\sim$}}}
\def\arcmin{\hbox{$^\prime$}}
\def\arcsec{\hbox{$^{\prime\prime}$}}
\def\arcdeg{\mbox{$^\circ$}}

\shorttitle{The Extreme SN~2008ha}
\shortauthors{Foley et~al.}

\begin{document}

 \title{SN~2008ha: An Extremely Low Luminosity and Extremely Low Energy Supernova}

\def\cfa{1}
\def\clay{2}
\def\berk{3}
\def\pitt{4}

\author{
{Ryan~J.~Foley}\altaffilmark{\cfa,\clay},
{Ryan~Chornock}\altaffilmark{\berk},
{Alexei~V.~Filippenko}\altaffilmark{\berk},
{Mohan~Ganeshalingam}\altaffilmark{\berk},
{Robert~P.~Kirshner}\altaffilmark{\cfa},
{Weidong~Li}\altaffilmark{\berk},
{S.~Bradley~Cenko}\altaffilmark{\berk},
{Pete~Challis}\altaffilmark{\cfa},
{Andrew~S.~Friedman}\altaffilmark{\cfa},
{Maryam~Modjaz}\altaffilmark{\berk},
{Jeffrey~M.~Silverman}\altaffilmark{\berk}, and
{W.~Michael~Wood-Vasey}\altaffilmark{\pitt}
}

\altaffiltext{\cfa}{
Harvard-Smithsonian Center for Astrophysics,
60 Garden Street, 
Cambridge, MA 02138.
}
\altaffiltext{\clay}{
Clay Fellow. Electronic address rfoley@cfa.harvard.edu .
}
\altaffiltext{\berk}{
Department of Astronomy,
University of California,
Berkeley, CA 94720-3411.
}
\altaffiltext{\pitt}{
Department of Physics and Astronomy,
University of Pittsburgh,
Pittsburgh, PA 15260.
}

\begin{abstract}
We present ultraviolet, optical, and near-infrared photometry as well
as optical spectra of the peculiar supernova (SN) 2008ha.  SN~2008ha
had a very low peak luminosity, reaching only $M_{V} = -14.2$~mag, and
low line velocities of only \about 2000~km~s$^{-1}$ near maximum
brightness, indicating a very small kinetic energy per unit mass of
ejecta.  Spectroscopically, SN~2008ha is a member of the
SN~2002cx-like class of SNe, a peculiar subclass of SNe~Ia; however,
SN~2008ha is the most extreme member, being significantly fainter and
having lower line velocities than the typical member, which is already
\about 2~mag fainter and has line velocities \about 5000~km~s$^{-1}$
smaller (near maximum brightness) than a normal SN~Ia.  SN~2008ha had
a remarkably short rise time of only \about 10~days, significantly
shorter than either SN~2002cx-like objects (\about 15~days) or normal
SNe~Ia (\about 19.5~days).  The bolometric light curve of SN~2008ha
indicates that SN~2008ha peaked at $L_{\rm peak} = (9.5 \pm 1.4)
\times 10^{40}$~ergs~s$^{-1}$, making SN~2008ha perhaps the least
luminous SN ever observed.  From its peak luminosity and rise time, we
infer that SN~2008ha generated $(3.0 \pm 0.9) \times 10^{-3}$
M$_{\sun}$ of $^{56}$Ni, had a kinetic energy of \about $2 \times
10^{48}$~ergs, and ejected 0.15 M$_{\sun}$ of material.  The host
galaxy of SN~2008ha has a luminosity, star-formation rate, and
metallicity similar to those of the Large Magellanic Cloud. We
classify three new (and one potential) members of the SN~2002cx-like
class, expanding the sample to 14 (and one potential) members.  The
host-galaxy morphology distribution of the class is consistent with
that of SNe~Ia, Ib, Ic, and II.  Several models for generating
low-luminosity SNe can explain the observations of SN~2008ha; however,
if a single model is to describe all SN~2002cx-like objects, either
electron capture in Ne-Mg white dwarfs causing a core collapse, or
deflagration of C-O white dwarfs with SN~2008ha being a partial
deflagration and not unbinding the progenitor star, are preferred.
The rate of SN~2008ha-like events is \about 10\% of the SN~Ia rate,
and in the upcoming era of transient surveys, several thousand similar
objects may be discovered, suggesting that SN~2008ha may be the tip of
a low-luminosity transient iceberg.
\end{abstract}

\keywords{supernovae---general, supernovae---individual(SN~1991bj,
SN~2002cx, SN~2004gw, SN~2006hn, SN~2007J, SN~2008ha),
galaxies---individual(UGC~12682)}


\defcitealias{Valenti09}{V09}

\section{Introduction}\label{s:intro}

Supernovae (SNe) are some of the most luminous and energetic events in
the Universe, having a luminosity of up to $M \approx -20$~mag at
peak, but SN~2008ha peaked at $M \approx -14$~mag and had spectra
which indicated a very low kinetic energy.  Although we do not know
the exact way in which most SNe explode or what the progenitors are
for most SNe, we have models which explain the vast majority of
stellar explosions; however, the standard SN models have difficulty
explaining the low luminosity and low ejecta velocity of SN~2008ha.
Most SNe are the result of either the density and temperature in the
core of a white dwarf (WD) increasing to the point of thermonuclear
runaway at or near the Chandrasekhar mass \citep[e.g.,][]{Hoyle60,
Nomoto84:w7} or a decrease in pressure support in the core of a
massive star leading to a gravitational collapse of the star that
results in a shock wave that destroys the star
\citep[e.g.,][]{Arnett89}.  Both of these scenarios produce (by
coincidence) SNe which have a few times $10^{50}$~ergs of kinetic
energy and peak at a few times $10^{9}$ L$_{\sun}$ (almost all of the
energy released from a thermonuclear SN is in the form of kinetic
energy, while most of the energy from a core-collapse SN is in the
form of neutrinos with only 1\% coupled to the baryonic matter).

SN~2008ha was discovered in UGC~12682 (an irregular galaxy with a
recession velocity of 1393~km~s$^{-1}$ and $(l, b) = (98.6^\circ,
-41.0^\circ)$ corresponding to a Virgo-infall corrected distance
modulus of $\mu = 31.64$~mag and, assuming $H_{0} =
73$~km~s$^{-1}$~Mpc$^{-1}$, $D = 21.3$~Mpc; all distances presented in
this paper are Virgo-infall corrected) on 2008 Nov.\ 7.17 (UT dates are
used throughout this paper) at mag 18.8 \citep{Puckett08}.  Our first
spectrum showed that it was a SN~Ia similar to SN~2002cx
\citep{Foley08:08ha}, a peculiar class of SNe \citep {Li03:02cx,
Jha06:02cx}.  SN~2008ha is similar in many ways to SN~2002cx, but
there are some differences.  SN~2008ha has an expansion velocity
\about 3000~km~s$^{-1}$ lower than that of SN~2002cx, which has
expansion velocities of \about 5000~km~s$^{-1}$ compared to \about
10,000~km~s$^{-1}$ for a normal SN~Ia.  SN~2008ha has a very small
peak absolute magnitude ($M \approx -14$~mag) while SN~2002cx has a
peak absolute magnitude of $M \approx -17$~mag, \about 2~mag below
that of normal SNe~Ia, but extending the relationship between decline
rate and luminosity \citep{Phillips93}, both have a decline rate that
is slower than expected from their luminosities.  Additionally,
SN~2002cx is distinguished from normal SNe~Ia (see
\citealt{Filippenko97} for a review of SN spectra) by having
spectra that resemble the high-luminosity SN~Ia 1991T
\citep{Filippenko92:91T, Phillips92} at early and intermediate phases,
except with significantly lower expansion velocities and having
late-time (\about 1 year after maximum) spectra which show
low-velocity \ion{Fe}{2} lines, few lines from intermediate-mass
elements, and no strong forbidden lines \citep{Jha06:02cx, Sahu08}.
Unfortunately, no maximum-light spectra were obtained and no very
late-time spectra have yet been obtained for SN~2008ha, so we cannot
compare SN~2008ha to SN~2002cx in these regards.

A recent study by \citet[hereafter V09]{Valenti09} has presented
unfiltered and $R$-band photometry and optical spectroscopy of
SN~2008ha, concluding that it is a low-luminosity and low-energy SN.
After comparing the spectra of SN~2008ha and a late-time spectrum of
SN~2005hk \citep[a SN~2002cx-like object;][]{Chornock06} to spectra of
low-luminosity SNe~II and examining the host-galaxy morphology
distribution of SN~2002cx-like objects, they conclude that SN~2008ha
and all SN~2002cx-like objects are core-collapse SNe.  They suggest
that these objects are the result of either electron capture in the
O-Mg-Ne core of 7--8 M$_{\sun}$ stars or the core collapse of $\gtrsim
30$ M$_{\sun}$ stars where material falls back onto a newly formed
neutron star, quickly creating a black hole which prevents a luminous
SN.

In this paper, we present additional observations, including
relatively early-time filtered observations covering the ultraviolet
(UV) through near-infrared (NIR), higher-resolution spectra, and a
relatively late-time spectrum.  Our observations indicate that
SN~2008ha may have been a core-collapse SN, but not of an asymptotic
giant branch (AGB) star.  Thermonuclear explosion models also fit the
observations.  Based on a sample of 14 SN~2002cx-like objects, we
conclude, in contrast to \citetalias{Valenti09}, that the host-galaxy
morphology distribution of SN~2002cx-like objects is consistent with
the population of SNe~Ia, particularly SN~1991T-like SNe.

We present and discuss our [$UVW1$]\ubvrijhk photometry in
\S~\ref{s:phot} and optical spectroscopy in \S~\ref{s:spec}.
In \S~\ref{s:energy}, we discuss the energetics of SN~2008ha.  We
examine the host galaxy of SN~2008ha and all SN~2002cx-like objects in
\S~\ref{s:hosts}.  In \S~\ref{s:model}, we present several
models which may explain our observations. Our results are summarized
in \S~\ref{s:disc}.


\section{Photometry}\label{s:phot}

\begin{figure}
\begin{center}
\epsscale{1.1}
\rotatebox{0}{
\plotone{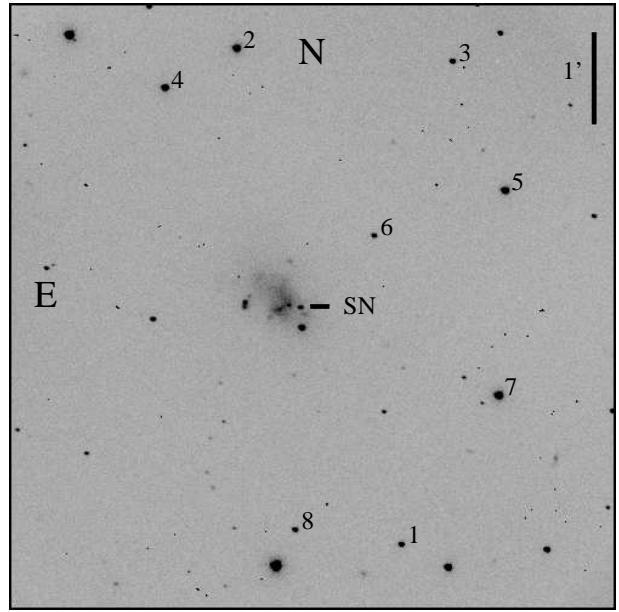}}
\caption{KAIT $R$-band image of SN~2008ha and its host galaxy,
UGC~12682.  The field of view (FOV) is 6.7\arcmin $\times$ 6.7\arcmin.
The SN and comparison stars are marked.  The labels for the comparison
stars correspond to the numbers in
Table~\ref{t:kait_stars}.}\label{f:finder}
\end{center}
\end{figure}

\begin{figure}
\begin{center}
\epsscale{2.1}
\rotatebox{90}{
\plotone{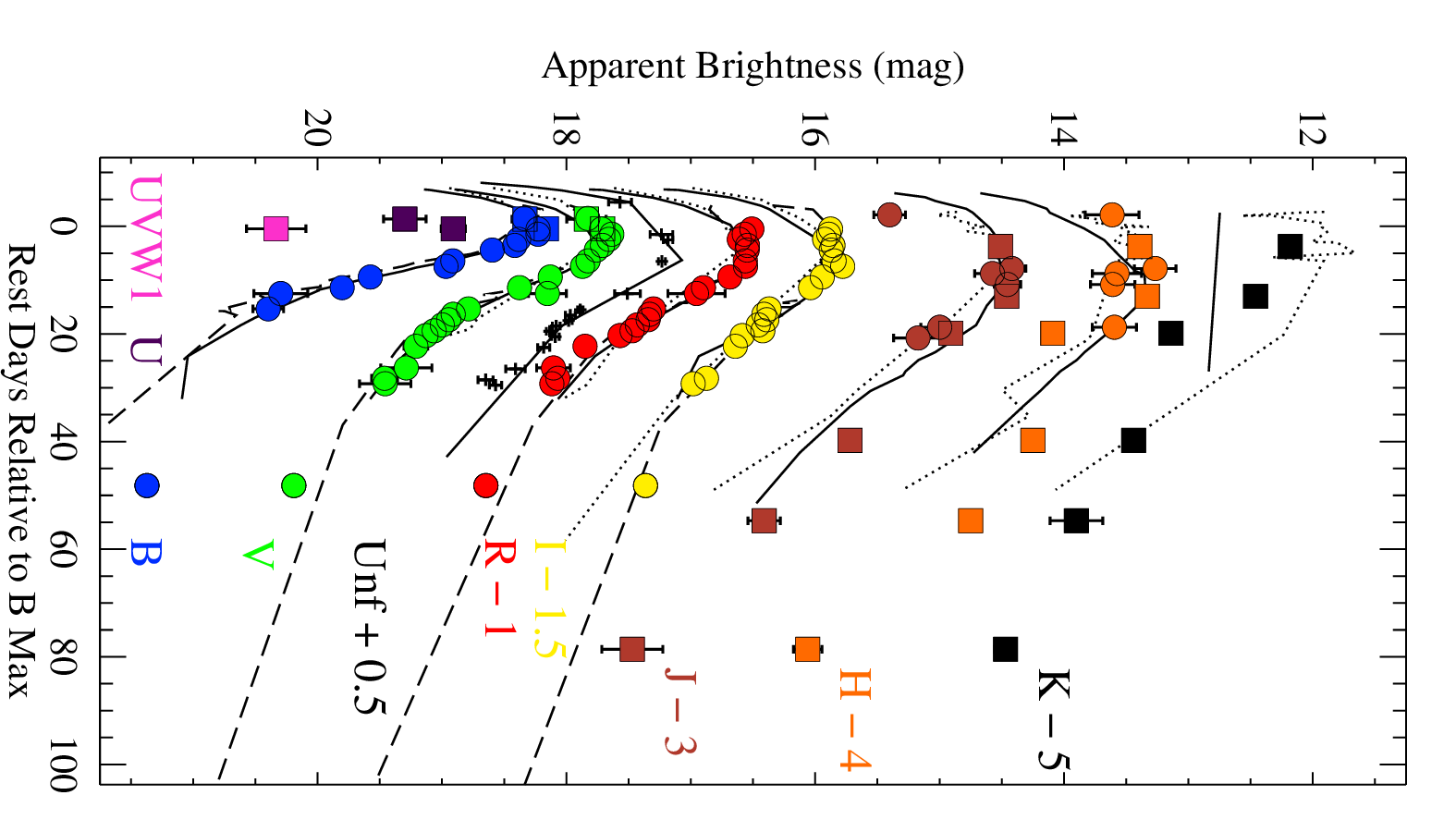}}
\caption{[$UVW1$]\ubvrijhk (circles and squares) and unfiltered (crosses, 
with the label ``Unf'') light curves of SN~2008ha.  Our unfiltered
magnitudes closely approximate the $R$ band.  The uncertainties for
most data points are smaller than the plotted symbols.  The comparison
light curves of SNe~2002cx (dashed lines), 2005hg (dotted lines), and
2005hk (solid lines) after stretching by 0.77, 0.70, and 0.73
(respectively) and offset to match the peak in each band, are also
plotted.  Our PANIC and NIRI \jhk points are plotted as squares to
distinguish them from the lower signal-to-noise ratio PAIRITEL points.
The {\it Swift} [$UVW1$]\ubv points are also plotted as squares.  We
see that the light curves of SN~2008ha are well matched by those of
the comparison light curves stretched by a factor of \about
0.7.}\label{f:lc}
\end{center}
\end{figure}

We have followed SN~2008ha photometrically with the 0.76~m Katzman
Automatic Imaging Telescope (KAIT; \citealt{Filippenko01}), the 1~m
Nickel 1~m, the 1.3~m Peters Automated Infrared Imaging Telescope
(PAIRITEL), the 6.5~m Magellan Baade telescope (with PANIC;
\citealt{Martini04}), and the 8-m Gemini-North telescope (with NIRI;
\citealt{Hodapp03}).  We also reduced public photometry from the {\it
Swift} satellite (with UVOT; \citealt{Roming05}.  A finding chart of
SN~2008ha, its host galaxy, and comparison stars is shown in
Figure~\ref{f:finder}.  Our [$UVW1$]\ubvrijhk light curves are
presented in Figure~\ref{f:lc}.

\subsection{Lick}

Accurate flux measurements of SN~2008ha require galaxy subtraction to
isolate the SN flux contribution.  A template for the KAIT unfiltered
data was constructed using previous observations of the field during
the Lick Observatory Supernova Search \citep{Filippenko01}.  \bvri
KAIT and Nickel templates were not available at the time of
publication.  To facilitate galaxy subtraction, templates were
manually constructed by taking a high-quality image and subtracting
the point-spread function (PSF) of the SN to produce a smooth galaxy
background.  Galaxy subtraction was not attempted with the Nickel data
which were taken under inferior conditions.  We performed differential
photometry on the SN and several local standard stars in the field
using the DAOPHOT package in IRAF\footnote{IRAF: the Image Reduction
and Analysis Facility is distributed by the National Optical Astronomy
Observatory, which is operated by the Association of Universities for
Research in Astronomy, Inc. (AURA) under cooperative agreement with
the National Science Foundation (NSF).} (see Table~\ref{t:kait_stars}
for a list of stars).  A single calibration was obtained under
photometric conditions using the Nickel telescope to determine the
zeropoint.  We adopted the statistical errors output by IRAF as our
final uncertainty.  Our KAIT and Nickel photometry can be found in
Table~\ref{t:kait}.

 \begin{deluxetable*}{lccccccc}
 \tablewidth{0pc}
 \tablecaption{Comparison Stars for SN~2008ha\label{t:kait_stars}}
 \tablehead{\colhead{Star} & \colhead{$\alpha$(J2000)} &\colhead{$\delta$(J2000)} &\colhead{$B$} (mag) & \colhead{$V$} (mag) & \colhead{$R$} (mag) & \colhead{$I$} (mag) & \colhead{$N_{\rm calib}$}}

\startdata

 SN & 23:34:52.69 & +18:13:35.4 & & & & & \\
  1 & 23:34:48.02 & +18:11:00.7 &   17.818(030) &   16.980(030) &   16.556(030) &   16.060(030) &  1\\
  2 & 23:34:55.69 & +18:16:26.0 &   16.353(030) &   15.407(030) &   14.903(030) &   14.433(030) &  1\\
  3 & 23:34:45.73 & +18:16:18.1 &   17.912(030) &   17.140(030) &   16.755(030) &   16.327(030) &  1\\
  4 & 23:34:58.99 & +18:15:59.9 &   17.600(030) &   16.123(030) &   15.188(030) &   14.267(030) &  1\\
  5 & 23:34:43.29 & +18:14:53.4 &   16.123(030) &   15.456(030) &   15.075(030) &   14.660(030) &  1\\
  6 & 23:34:49.33 & +18:14:23.2 &   18.817(030) &   17.670(030) &   16.976(030) &   16.394(030) &  1\\
  7 & 23:34:43.57 & +18:12:38.7 &   16.831(030) &   15.425(030) &   14.519(030) &   13.690(030) &  1\\
  8 & 23:34:52.92 & +18:11:10.0 &   17.719(030) &   17.072(030) &   16.755(030) &   16.362(030) &  1

\enddata

\tablecomments{Uncertainties are given in parentheses.}

\end{deluxetable*}

 \begin{deluxetable*}{lcccccc}
 \tablewidth{0pc}
 \tablecaption{Lick Photometry of SN~2008ha\label{t:kait}}
 \tablehead{\colhead{JD} & \colhead{$B$ (mag)} & \colhead{$V$ (mag)} & \colhead{$R$ (mag)} & \colhead{$I$ (mag)} & \colhead{Unfiltered (mag)} & \colhead{Telescope}}
 \startdata
2454764.80 & \nodata       & \nodata       & \nodata       & \nodata       & $>19.5$        & KAIT \\
2454778.69 & \nodata       & \nodata       & \nodata       & \nodata       & 18.069 (092)   & KAIT \\
2454781.76 & 18.340  (084) & 17.828  (037) & \nodata       & \nodata       & \nodata        & KAIT \\
2454783.74 & 18.229  (062) & 17.718  (042) & 17.509  (041) & 17.377  (054) & \nodata        & KAIT \\
2454784.64 & \nodata       & \nodata       & \nodata       & \nodata       & 17.736 (091)   & KAIT \\
2454784.71 & 18.230  (030) & 17.635  (030) & 17.570  (030) & 17.392  (030) & \nodata        & Nickel \\
2454785.67 & 18.385  (030) & 17.660  (027) & 17.610  (023) & 17.425  (034) & 17.683 (038)   & KAIT \\
2454786.80 & 18.415  (030) & 17.710  (030) & 17.544  (030) & 17.358  (030) & \nodata        & Nickel \\
2454787.67 & 18.596  (030) & 17.762  (030) & 17.552  (030) & 17.376  (030) & \nodata        & KAIT \\
2454789.68 & 18.904  (030) & 17.827  (014) & 17.573  (011) & 17.353  (030) & 17.732 (028)   & KAIT \\
2454790.71 & 18.969  (030) & 17.871  (030) & 17.560  (030) & 17.278  (030) & \nodata        & Nickel \\
2454792.70 & 19.563  (054) & 18.127  (022) & 17.689  (016) & 17.443  (030) & \nodata        & KAIT \\
2454794.68 & 19.789  (072) & 18.377  (032) & 17.904  (025) & 17.541  (037) & \nodata        & KAIT \\
2454795.78 & 20.297  (216) & \nodata       & \nodata       & \nodata       & 18.009 (103)   & KAIT \\
2454798.65 & 20.390  (123) & 18.791  (033) & 18.308  (022) & 17.876  (039) & 18.388 (022)   & KAIT \\
2454799.66 & \nodata       & 18.921  (046) & 18.343  (029) & 17.919  (050) & 18.468 (040)   & KAIT \\
2454800.67 & \nodata       & 18.942  (035) & 18.347  (029) & 17.892  (033) & 18.482 (027)   & KAIT \\
2454801.64 & \nodata       & 19.004  (044) & 18.438  (026) & 17.964  (036) & 18.589 (031)   & KAIT \\
2454802.74 & \nodata       & 19.064  (082) & 18.472  (037) & 17.921  (030) & 18.635 (026)   & KAIT \\
2454803.63 & \nodata       & 19.130  (054) & 18.569  (031) & 18.087  (047) & 18.589 (030)   & KAIT \\
2454805.63 & \nodata       & 19.211  (095) & 18.851  (055) & 18.142  (056) & 18.682 (048)   & KAIT \\
2454809.67 & \nodata       & 19.288  (206) & 19.104  (135) & \nodata       & 18.909 (078)   & KAIT \\
2454811.61 & \nodata       & 19.462  (103) & 19.069  (073) & 18.374  (067) & 19.150 (060)   & KAIT \\
2454812.62 & \nodata       & 19.458  (207) & 19.118  (065) & 18.482  (077) & 19.070 (049)   & KAIT \\
2454831.63 & 21.372  (030) & 20.191  (030) & 19.647  (030) & 18.864  (030) & \nodata        & Nickel

\enddata

\tablecomments{Uncertainties are given in parentheses.}

\end{deluxetable*}

\subsection{PAIRITEL}

We obtained NIR (\jh\!\!) photometry with the refurbished and fully
automated 1.3-m Peters Automated Infrared Telescope
(PAIRITEL)\footnote{See http://www.pairitel.org/ .} located at FLWO.
$J$- and $H$-band (1.2 and 1.6~$\mu$m; \citealt{Cohen03}) images were
acquired simultaneously with the three NICMOS3 arrays with individual
exposure times of 7.8~s. The PAIRITEL reduction pipeline software
\citep{Bloom06:PAIRITEL} estimates the sky background from a
star-masked median stack of the SN raw images. After sky subtraction,
it then cross correlates, stacks, and subsamples the processed images
in order to produce the final image with an effective scale of
1\arcsec~pixel$^{-1}$ and an effective FOV of 10\arcmin $\times$
10\arcmin.

For the NIR photometry, we used the image-analysis pipeline of the
ESSENCE and SuperMACHO projects and performed difference photometry
(following \citealt{Wood-Vasey08}) using the image-subtraction
algorithm of \citet{Alard00}.  We present our PAIRITEL \jh light
curves in Table~\ref{t:pairitel}.

%
%
%
%

\begin{deluxetable*}{lcc}
\tablewidth{0pc}
\tablecaption{PAIRITEL Photometry of SN~2008ha\label{t:pairitel}}
\tablehead{\colhead{JD} & \colhead{$J$} (mag) & \colhead{$H$} (mag)}
\startdata

2454781.09 & 18.401 (254) & 17.613 (432) \\
2454791.12 & 17.418 (223) & 17.266 (334) \\
2454792.05 & 17.574 (287) & 17.574 (396) \\
2454794.09 & 17.455 (212) & 17.608 (358) \\
2454802.08 & 17.999 (303) & 17.595 (356) \\
2454804.09 & 18.173 (392) & \nodata      

\enddata

\tablecomments{Uncertainties are given in parentheses.}

\end{deluxetable*}

\subsection{PANIC and NIRI}

Observations in \jhk were obtained using Magellan PANIC in classically
scheduled mode and Gemini NIRI as part of the director's discretionary
time (program ID: GN-2008B-DD-6).

The NIRI data were reduced using the \texttt{noao:gemini:niri} IRAF
package and its bad-pixel masks. Sky flats were generated using the
7--15 different dither positions that comprised each sequence.  After
the standard NIRI reduction, we used IRAF \texttt{imcoadd} to combine
the dithers in a mosaic stack.  Simple aperture photometry
(0.6\arcsec\ radius) then gave the fluxes for SN~2008ha and the 1--2
2MASS stars (\about 15~mag) in the resulting mosaics.  We estimate a
systematic uncertainty of 0.03~mag from using two stars and 0.05~mag
from using just one 2MASS star.

The PANIC data were reduced using the standard IRAF PANIC reduction
software available at the LCO website.  We constructed a bad-pixel
mask from on/off pairs of dark/lamp frames and then used the sky
dithers for our flat field.  The same 2MASS stars were used as
calibration sources as for the NIRI data.

Both the NIRI and PANIC photometry was done with local sky subtraction
in a circular annulus.  With the good resolution of the Magellan and
Gemini images (\about 0.7\arcsec) the SN stood out clearly from the
galaxy background, and we estimate that the residual uncertainty in
the subtraction of the local sky/galaxy background is 0.01~mag.  Our
light curves are presented in Figure~\ref{f:lc} and
Table~\ref{t:niri}.

\begin{deluxetable*}{lllllllllll}

\tablecolumns{11}
\tablecaption{PANIC and NIRI Observations of SN~2008ha\label{t:niri}}
\tablehead{
\colhead{JD}   & 
  \colhead{$J$} (mag) & \colhead{(stat)}  & \colhead{(cal)\tablenotemark{a}}   & 
  \colhead{$H$} (mag) & \colhead{(stat)}  & \colhead{(cal)\tablenotemark{a}}   & 
  \colhead{$K$} (mag) & \colhead{(stat)}  & \colhead{(cal)\tablenotemark{a}}   &
  \colhead{Detector}
}
\startdata
2454787.0    & 17.51  & 0.01   & 0.05  & 17.39  & 0.02   & 0.05  & 17.18  & 0.04   & 0.05  & PANIC \\
2454796.3    & 17.46  & 0.03   & 0.05  & 17.33  & 0.03   & 0.03  & 17.46  & 0.05   & 0.03  &  NIRI \\
2454803.2    & 17.91  & 0.01   & 0.05  & 18.09  & 0.02   & 0.05  & 18.14  & 0.04   & 0.05  &  NIRI \\
2454823.2    & 18.72  & 0.05   & 0.05  & 18.25  & 0.05   & 0.03  & 18.44  & 0.05   & 0.03  &  NIRI \\
2454838.2    & 19.41  & 0.12   & 0.05  & 18.75  & 0.03   & 0.03  & 18.90  & 0.21   & 0.03  &  NIRI \\
2454862.2    & 20.47  & 0.24   & 0.05  & 20.06  & 0.11   & 0.03  & 19.47  & 0.08   & 0.03  &  NIRI
\enddata

\tablecomments{Observations calibrated to 2MASS stars.  No airmass or
color corrections applied.}

\tablenotetext{a}{Calibration uncertainties are 0.03 for one star and
0.05 for two as a rough summation of system and stability
uncertainties.}

\end{deluxetable*}

\subsection{{\it Swift}}

The {\it Swift} team initiated target of opportunity observations of
SN~2008ha with the Ultraviolet/Optical Telescope (UVOT;
\citealt{Roming05}) and the X-ray Telescope (XRT; \citealt{Burrows05})
on board the {\it Swift} gamma-ray burst satellite \citep{Gehrels04}
beginning on 2008 Nov. 11.  A second, deeper epoch was obtained two
days later near the optical peak, and a final set of reference images
was obtained (at our request) on 2009 Feb. 3.  We performed digital
image subtraction on all of the UVOT data using the final epoch as a
template to remove host-galaxy contamination with the ISIS software
package \citep{Alard00}.  The $U$-, $B$-, and $V$-band data were then
reduced using the photometric calibration technique described by
\citet{Li06}, while the UV data were photometered using the zeropoints
from \citet{Poole08}.  The results of our UVOT analysis are displayed
in Table~\ref{t:swift}.

The XRT data were reduced using standard {\it Swift} software analysis
tools.  No X-ray source is detected at the location of SN~2008ha in
any of our three epochs of observations.  Assuming a power-law
spectrum with photon index $\Gamma = 2.0$, we place the following
limits on the $0.3$--$10$ keV X-ray flux from SN~2008ha: 2008 Nov.\
11.32, $F_{X} < 9 \times 10^{-14}$~ergs~cm$^{-2}$~s$^{-1}$; 2008 Nov.\
13.19, $F_{X} < 5 \times 10^{-14}$~ergs~cm$^{-2}$~s$^{-1}$; 2009 Feb.\
3.04, $F_{X} < 7 \times 10^{-14}$~ergs~cm$^{-2}$~s$^{-1}$.  Using our
adopted distance modulus, these fluxes correspond to isotropic
luminosities of $L_{X} < 5 \times 10^{39}$~ergs~s$^{-1}$, $L_{X} < 3
\times 10^{39}$~ergs~s$^{-1}$, and $L_{X} < 4 \times
10^{39}$~ergs~s$^{-1}$, respectively.  Several SNe~Ia have been
measured to have no X-ray emission at early times with limits an order
of magnitude deeper \citep{Hughes07}, while SNe~Ic typically have
X-ray luminosities of $10^{37}$--$10^{39}$~ergs~s$^{-1}$
\citep{Chevalier06}; therefore, our X-ray limits are not particularly
constraining.

 \begin{deluxetable*}{lcccccc}
 \tablewidth{0pc}
 \tablecaption{{\it Swift} Photometry of SN~2008ha\label{t:swift}}
 \tablehead{\colhead{JD} & \colhead{$UVW2$} (mag) & \colhead{$UVM2$} (mag) & \colhead{$UVW1$} (mag) & \colhead{$U$} (mag) & \colhead{$B$} (mag) & \colhead{$V$} (mag)}
 \startdata
2454781.9 & \nodata    & \nodata    & $> 20.179$   & 19.299 (171) & 18.333 (104) & 17.839 (158) \\
2454783.7 & $> 20.072$ & $> 19.647$ & 20.333 (239) & 18.909 (096) & 18.161 (051) & 17.709 (067)

\enddata
\tablecomments{Uncertainties are given in parentheses.}

\end{deluxetable*}

\subsection{Light Curves}

By fitting a polynomial to the light-curve peaks, we are able to
measure the time of maximum brightness and peak brightness in
\bvrijhk for SN~2008ha.  We find that SN~2008ha peaked in the $B$ band
on JD $2,454,783.23 \pm 0.16$ at $18.23 \pm 0.01$~mag.  SN~2008ha also
peaked at $V = 17.68 \pm 0.01$~mag.  From our medium-resolution MagE
spectra (see \S~\ref{s:spec}), we can place a limit of $< 0.02$~\AA\
for the equivalent with of Na~D from the host galaxy, indicating
minimal host-galaxy extinction.  Using the distance modulus for
UGC~12682, and correcting for Milky Way extinction of $A_{V} =
0.25$~mag \citep[][consistent with our measured equivalent width of
the Na~D lines corresponding to Milky Way absorption]{Schlegel98}, we
find that SN~2008ha peaked at $M_{V} = -14.21 \pm 0.15$~mag, about
5~mag below the peak absolute magnitude of normal SNe~Ia and 2.9~mag
fainter than the least luminous known SN~Ia (SN~2007ax peaked at
$M_{V} = -17.1$~mag; \citealt{Kasliwal08}).  The maximum-light
characteristics of SN~2008ha for \bvrijhk can be found in
Table~\ref{t:photdata}.

\begin{deluxetable*}{lrrrrrrr}
\tabletypesize{\scriptsize}
\tablewidth{0pt}
\tablecaption{Photometric Information for SN~2008ha\label{t:photdata}}
\tablehead{
\colhead{Filter} &
\colhead{$B$ (mag)} &
\colhead{$V$ (mag)} &
\colhead{$R$ (mag)} &
\colhead{$I$ (mag)} &
\colhead{$J$ (mag)} &
\colhead{$H$ (mag)} &
\colhead{$K$ (mag)}}

\startdata

JD of max $-$ 2,454,000   & $783.23 \pm 0.16$ & $785.24 \pm 0.30$ & $787.30 \pm 0.18$ & $787.95 \pm 0.28$ & $791.22 \pm 0.81$ & $790.7 \pm 3.9$   & $788.9 \pm 5.8$   \\
Mag at max            & $18.23 \pm 0.01$  & $17.68 \pm 0.01$  & $17.54 \pm 0.01$  & $17.36 \pm 0.01$  & $17.46 \pm 0.02$  & $17.12 \pm 0.27$  & $17.10 \pm 0.55$  \\
Peak abs.\ mag        & $-13.74 \pm 0.15$ & $-14.21 \pm 0.15$ & $-14.31 \pm 0.15$ & $-14.43 \pm 0.15$ & $-14.24 \pm 0.15$ & $-14.56 \pm 0.30$ & $-14.57 \pm 0.57$ \\
$\Delta m_{15}$ (mag) & $2.17 \pm 0.02$   & $1.22 \pm 0.03$   & $0.97 \pm 0.02$   & $0.65 \pm 0.02$   & $0.51 \pm 0.04$   & $0.82 \pm 0.42$   & $1.05 \pm 0.54$
 
\enddata

\end{deluxetable*}

\citetalias{Valenti09} presented $R$-band and unfiltered photometry of
SN~2008ha.  Their light curves are broadly consistent with ours.
Their first filtered data point occurred on JD 2,454,798.46 with $R =
18.36 \pm 0.12$~mag.  This is consistent within the uncertainties of
our measurement 0.19~days later at $R = 18.308 \pm 0.022$~mag.  Their
derived peak magnitude is 0.3~mag brighter than our measurement, but
with the same time of maximum.  Considering that their first filtered
observations occur 12~days after our measured date of maximum, the
disagreement between our measurement and their extrapolation is not
surprising.  They adopted a distance modulus of $\mu = 31.55$~mag,
0.09~mag smaller than our adopted value.  As a result of these two
differences, they measured a peak absolute magnitude that is 0.2~mag
brighter than what we have measured.  Their late-time $R$-band points
are systematically fainter than our late-time points, probably from
undersubtraction (for our photometry) or oversubtraction (for their
photometry) of the host-galaxy light.  This discrepancy should be
reduced after late-time templates are obtained to perform template
subtractions.

Comparing our light curves to those of SN~2002cx \citep{Li03:02cx}, SN
2005hk (another object similar to SN~2002cx; \citealt{Phillips07}),
and SN 2005hg (a normal SN~Ib; M. Modjaz, 2009, private
communication), we find that our \bvrijhk light curves are decline
quickly, but are very similar to all three SNe if we compress the
light curves in time (``stretching'' the light curves by \about 0.7
for all three; see Figure~\ref{f:lc}).  The date of $B$ maximum found
from matching SN~2008ha to SNe~2002cx and 2005hk is JD $2,454,783.6
\pm 0.1$, very similar what we found from polynomial fitting.

Our strongest limit on the rise time of SN~2008ha comes from a
nondetection in a KAIT search image on 2008 Oct.\ 25 ($m_{\rm unf} >
19.5$~mag), 18.4~days before $B$-band maximum. However, if we assume
that SNe~2005hk and 2008ha have similar compositions, opacities, and
temperatures (which are reasonable considering their spectral
similarities; see \S~\ref{s:spec}), we can estimate the rise time of
SN~2008ha by scaling the light curve of SN~2005hk (which has a
well-constrained rise time) to match that of SN~2008ha.  SN~2005hk has
a rise time of 15~days \citep{Phillips07}, and with a scaling factor
of 0.7, we determine that the rise time of SN~2008ha is \about $0.7
t_{r} ({\rm SN~2005hk}) \approx 10$~days.

Finally, if we assume that SN~2008ha is a homologously expanding
object, then we can use an analytic equation to solve for the rise
time \citep{Riess99:risetime}.  This assumption only holds true at
very early times.  Using our unfiltered measurements, which are the
earliest data, and the discovery measurement from \citet{Puckett08}
with no correction to match their unfiltered measurements to ours, we
determine that the rise time in the unfiltered band is 11~days (if we
only use the earliest 2 data points) to 18~days (if we use the
earliest 3 data points).  From our polynomial fits, we see that there
is a 4~day lag between $B$ maximum and unfiltered maximum; applying
this lag to the rise time, we find that SN~2008ha has a $B$-band rise
time of 7--14~days, consistent with the value found above by matching
light curves.  Our first value (7~days) is measured with only two data
points, leaving zero degrees of freedom in our fit.  Our second value
(14~days) is measured from a fit that includes a light-curve point
very close to maximum brightness, so the derived rise time should be
considered a relatively strong upper limit.

In Figure~\ref{f:lc}, we see that the light curves of both
SN~2002cx-like objects (represented by SNe~2002cx and 2005hk) and
SNe~Ib are similar near maximum light.  The offsets in absolute
magnitude and stretch factors necessary for each comparison SN to
match SN~2008ha are listed in Table~\ref{t:lcparam}.  SN~2002ha does
not have the second maximum in the NIR bands that is common to SNe~Ia;
however, there appears to be a shoulder in $H\!K$ around 20--40~days
after $B$-band maximum.  SN~2005hk did not have a similar shoulder in
its $H$-band light curve (the $K$-band coverage for SN~2005hk is only
two points, and does not allow a good comparison to SN~2008ha).  The
extinction-corrected colors at maximum of SNe~2002cx and 2005hk are
more similar to SN~2008ha than those of SN~2005hg ($\sigma = 0.20$,
0.30, and 0.16~mag for SNe~2002cx, 2005hg, and 2005hk, respectively).

\begin{deluxetable*}{llcccccccc}
\tabletypesize{\scriptsize}
\tablewidth{0pt}
\tablecaption{Light-Curve Fit Parameters\label{t:lcparam}}
\tablehead{
\multicolumn{3}{c}{ } &
\multicolumn{7}{c}{Abs.\ Magnitude Offset} \\
\colhead{SN Name} &
\colhead{SN Type} &
\colhead{Stretch} &
\colhead{$B$} &
\colhead{$V$} &
\colhead{$R$} &
\colhead{$I$} &
\colhead{$J$} &
\colhead{$H$} &
\colhead{$K$}}

\startdata

SN~2002cx & 02cx-like  & 0.77 & 3.7     & 3.2 & 3.3 & 3.2 & \nodata & \nodata & \nodata \\
SN~2005hk & 02cx-like  & 0.73 & 3.9     & 3.7 & 3.7 & 3.7 & 3.4     & 3.6     & 3.3 \\
SN~2005hg & Ib         & 0.70 & 3.4     & 3.3 & 3.4 & 3.3 & 4.0     & 4.0     & 3.5

\enddata

\end{deluxetable*}

\subsection{Spectral Energy Distribution}

\begin{figure}
\begin{center}
\epsscale{0.88}
\rotatebox{90}{
\plotone{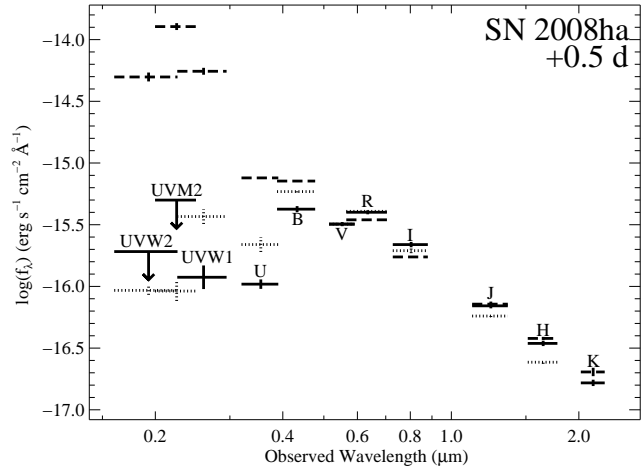}}
\caption{SED of SN~2008ha (solid crosses) constructed from UVOIR
broad-band filters on $t = 0.5$~days (for the UV and optical, the NIR
points are from $t = 3.8$~d).  Each filter is labeled.  The UVW2 and
UVM2 filters yield only upper limits.  We also plot near-maximum SEDs
of SNe~2005cs (dashed crosses) and 2005hk (dotted crosses) shifted to
match the $V$-band flux of SN~2008ha.  SN~2008ha has a less UV flux
than SN~2005hk and significantly less UV flux than
SN~2005cs.}\label{f:sed}
\end{center}
\end{figure}

On 2008 Nov.\ 13 (JD 2,454,783.7), 0.5~days after $B$ maximum, we have
concurrent measurements in the UV (from {\it Swift} using the
zeropoints of \citealt{Brown08}) and optical (from KAIT).  We also
have NIR measurements (from Magellan) 3.3~days later.  From
Figure~\ref{f:lc}, we see that the \jhk bands are relatively constant
between these epochs.  A spectral energy distribution (SED) for
SN~2008ha covering $\lambda < 0.2$~$\mu$m to $\lambda > 2.2$~$\mu$m,
is shown in Figure~\ref{f:sed}.  We see that the SED has a flat
$f_{\lambda}$ spectrum in the optical and a remarkable drop in flux in
the UV.

We can compare the SED of SN~2008ha to that of other SNe that have
UV-optical-IR (UVOIR) coverage near maximum light.  In
Figure~\ref{f:sed}, we plot the near-maximum light SEDs of SN~2005cs
\citep{Pastorello09} and SN~2005hk \citep{Phillips07, Brown08}.  All
of these SNe have relatively similar optical and NIR colors, but their
UV fluxes differ significantly.  SN~2005cs has significant UV flux,
while SNe~2005hk and 2008ha have line blanketing in their photosphere
causing a depression of UV flux.


\section{Spectroscopy}\label{s:spec}

We have obtained several low and medium-resolution spectra of
SN~2008ha with the FAST spectrograph \citep{Fabricant98} on the FLWO
1.5~m telescope, the Kast double spectrograph \citep{Miller93} on the
Shane 3~m telescope at Lick Observatory, the MagE spectrograph
\citep{Marshall08} on the Magellan Clay 6.5~m telescope, the Blue
Channel spectrograph \citep{Schmidt89} on the 6.5~m MMT telescope, and
the Low Resolution Imaging Spectrometer \citep[LRIS;][]{Oke95} on the
10~m Keck~I telescope.

\begin{figure*}
\begin{center}
\epsscale{1.45}
\rotatebox{90}{
\plotone{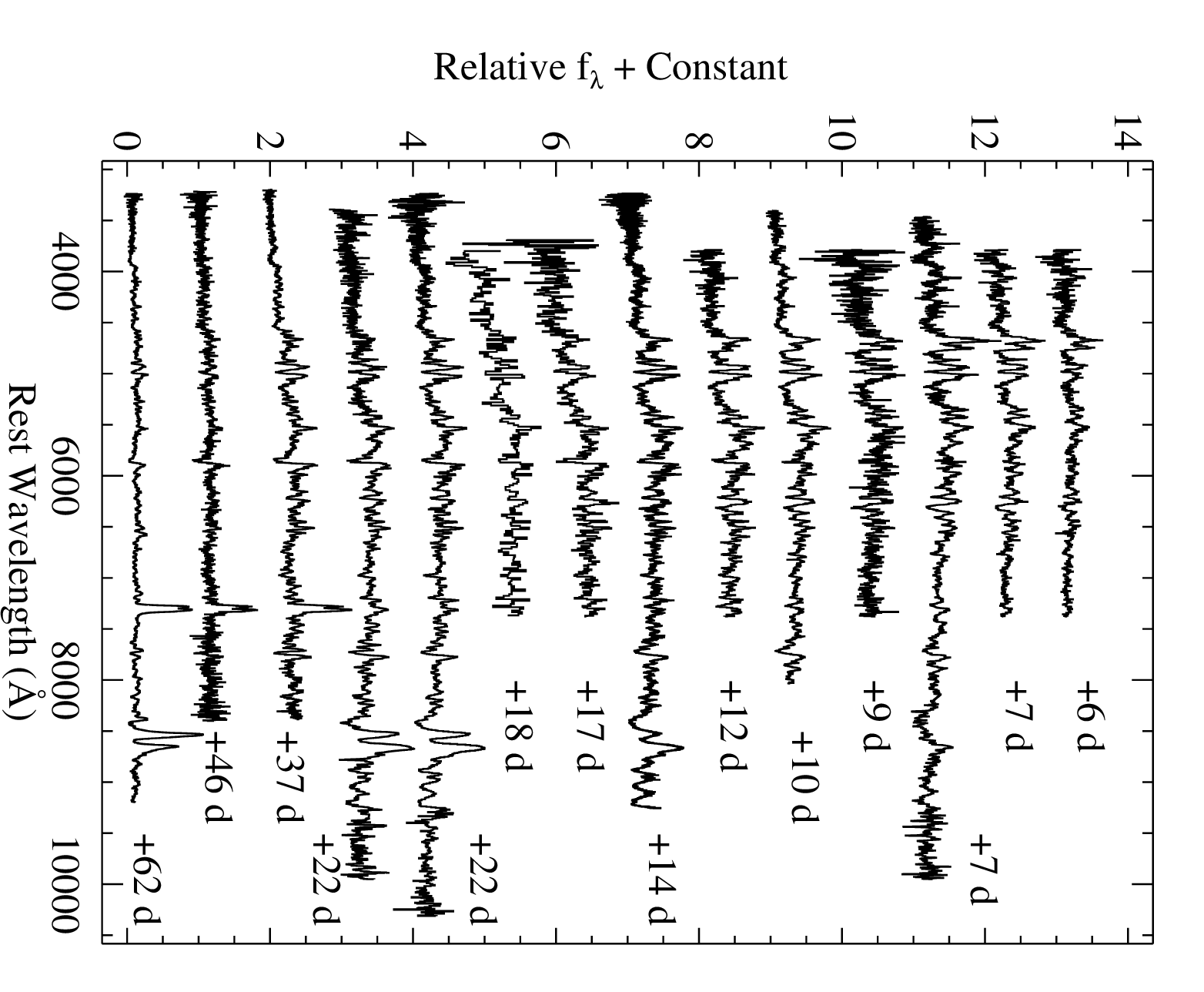}}
\caption{Optical spectra of SN~2008ha.  The spectra are denoted by
their phase relative to maximum brightness in the $B$ band.  For
clarity, we have interpolated over nebular emission lines from the
host galaxy.  Note that most spectra have a relatively high
signal-to-noise ratio, but the low line velocities make the spectra
appear noisier than they really are.  See Figure~\ref{f:synow_08ha}
for a better representation of the quality of our data.}\label{f:spec}
\end{center}
\end{figure*}

\begin{deluxetable*}{lllrl}
\tabletypesize{\scriptsize}
\tablewidth{0pt}
\tablecaption{Log of Spectral Observations\label{t:spec}}
\tablehead{
\colhead{} &
\colhead{} &
\colhead{Telescope /} &
\colhead{Exposure} &
\colhead{} \\
\colhead{Phase\tablenotemark{a}} &
\colhead{UT Date} &
\colhead{Instrument} &
\colhead{(s)} &
\colhead{Observer\tablenotemark{b}}}

\startdata

 6.5 & 2008 Nov.\ 19.2 & FLWO/FAST        & 1800                 & NW \\
 7.5 & 2008 Nov.\ 20.2 & FLWO/FAST        & $2 \times 1800$      & NW \\
 7.5 & 2008 Nov.\ 20.2 & Lick/Kast        & 2400                 & CG, XW \\
 9.4 & 2008 Nov.\ 22.1 & FLWO/FAST        & $2 \times 1800$      & NW \\
10.5 & 2008 Nov.\ 23.2 & Lick/Kast        & 1800                 & BC, MM \\
12.4 & 2008 Nov.\ 25.1 & FLWO/FAST        & $2 \times 1800$      & AV \\
14.3 & 2008 Nov.\ 27.0 & Clay/MagE        & 1800                 & JB \\
17.5 & 2008 Nov.\ 30.2 & FLWO/FAST        & $2 \times 1800$      & WP \\
18.4 & 2008 Dec.\ 1.1  & FLWO/FAST        & $2 \times 1800$      & WP \\
22.3 & 2008 Dec.\ 5.0  & Clay/MagE        & 200, $2 \times 900$  & AR, WH \\
22.5 & 2008 Dec.\ 5.2  & Lick/Kast        & $3 \times 1800$      & FS, XW \\
37.4 & 2008 Dec.\ 20.1 & MMT/Blue Channel & 600, $3 \times 1800$ & PC \\
46.4 & 2008 Dec.\ 29.1 & MMT/Blue Channel & $3 \times 1350$      & DS \\
62.5 & 2009 Jan.\ 14.2 & Keck/LRIS        & $2 \times 1800$      & RC

\enddata

\tablenotetext{a}{Days since $B$ maximum, 2008 Nov.\ 12.7 (JD 2,454,783.2).}

\tablenotetext{b}{AR = A.\ Rest, AV = A.\ Vaz, BC = B.\ Cenko, CG =
C.\ Griffith, DS = D.\ Sand, FS = F.\ Serduke, JB = J.\ Bochanski, MM
= M.\ Modjaz, NW = N.\ Wright, PC = P.\ Challis, RC = R.\ Chornock, WH
= W.\ High, WP = W.\ Peters, XW = X.\ Wang.}

\end{deluxetable*}

Standard CCD processing and spectrum extraction were accomplished with
IRAF.  The data were extracted using the optimal algorithm of
\citet{Horne86}.  Low-order polynomial fits to calibration-lamp
spectra were used to establish the wavelength scale, and small
adjustments derived from night-sky lines in the object frames were
applied.  For the MagE spectra, the sky was subtracted from the images
using the method described by \citet{Kelson03}.  We employed our own
IDL routines to flux calibrate the data and remove telluric lines
using the well-exposed continua of the spectrophotometric standards
\citep{Wade88, Foley03}.

In this Section, we will present spectra from the literature,
including a spectrum from Perets et~al.\ (in prep.), and previously
unpublished spectra of SNe~2005E, 2005cs, and 2007J.  Our spectra of
SNe~2005E and 2005cs were obtained with LRIS mounted on Keck~I on 2005
Mar.\ 11.3 and 2006 Apr.\ 27.5, respectively.  Our spectra of SN~2007J
were obtained with LRIS on 2007 Jan.\ 21.4 and 2007 Mar.\ 18.3 (see
the Appendix).

\subsection{Spectroscopic Properties of SN~2008ha}

In Figure~\ref{f:spec}, we present several spectra of SN~2008ha from
the classifying spectrum \citep{Foley08:08ha} to a spectrum obtained
in 2009 January (corresponding to phases of $t = +6.5$ to 62.5~days
after maximum brightness in the $B$ band).  A journal of our
observations can be found in Table~\ref{t:spec}.  Our earliest spectra
show the distinct characteristics of the SN~2002cx-like class of
objects.  The spectra exhibit low-velocity lines of intermediate-mass
and Fe-group elements.  Examining relatively unblended lines such as
\ion{O}{1} and Na~D, we measure typical ejecta velocities of \about
2000~km~s$^{-1}$.  \citetalias{Valenti09} found a higher ejecta
velocity of 2300~km~s$^{-1}$.  Considering the relatively large
systematic uncertainties, we believe that our measurements are
consistent with those found by
\citetalias{Valenti09}.

At late times, the spectra evolve in a manner similar to that of
SN~2002cx, except that SN~2008ha has exceptionally strong
[\ion{Ca}{2}] $\lambda \lambda 7291$, 7323 and \ion{Ca}{2} NIR triplet
($\lambda \lambda 8498$, 8542, 8662) emission lines.  The lack of H
lines at any epoch indicates that the progenitor of SN~2008ha did not
have a massive H envelope.  Since it is difficult to excite the
optical lines of He, the absence of He lines in our spectra does not
necessarily mean that there was no He in the ejecta.

\begin{figure}
\begin{center}
\epsscale{0.88}
\rotatebox{90}{
\plotone{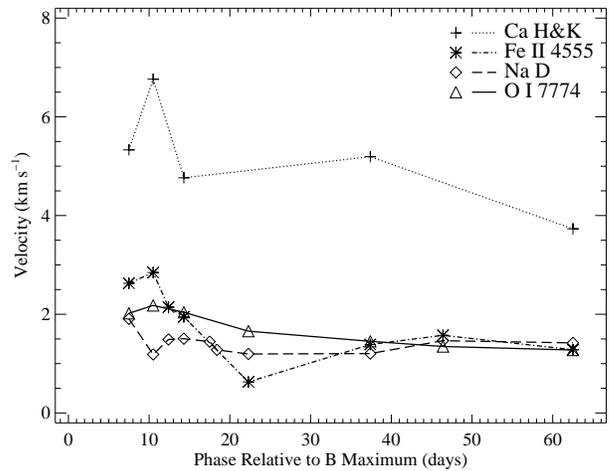}}
\caption{Velocity evolution of the Ca H\&K (dotted line and crosses),
\ion{Fe}{2} $\lambda 4555$ (dot-dashed line and asterisks), Na~D
(dashed line and diamonds), and \ion{O}{1} $\lambda 7774$ (solid line
and triangles) lines as a function of phase for SN~2008ha.  The Ca
H\&K doublet may be contaminated by other features, affecting our
velocity measurements.}\label{f:vel}
\end{center}
\end{figure}

Tracking the minima of the absorption for several P-Cygni lines, we
can measure the distribution of elements within the ejecta.  In
Figure~\ref{f:vel}, we show the velocity evolution for four lines (Ca
H\&K, \ion{Fe}{2} $\lambda 4555$, Na~D, and \ion{O}{1} $\lambda
7774$).  We see only a modest decrease in the velocities of Na~D and
\ion{O}{1} $\lambda 7774$ of only \about 500 and \about
800~km~s$^{-1}$ over a 60-day period, respectively. \ion{Fe}{2}
$\lambda 4555$, on the other hand, decreased by \about
1300~km~s$^{-1}$ over the same period, with all three lines having the
same velocity in our last spectrum.  Meanwhile, Ca H\&K has a much
higher velocity than the other lines.  For our $+8$~day spectrum we
find that the \ion{Ca}{2} NIR triplet has a velocity similar to that
of Ca H\&K, so we do not believe that the higher velocity is the
result of blending of the feature with other lines.  \ion{Ca}{2} has a
very large optical depth, so we may be seeing the outermost layers of
the ejecta moving at larger velocities for Ca H\&K.  All lines
measured are multiplets and at the small line widths, the individual
components can create systematic errors in measuring the minima of the
features.  The Ca H\&K, \ion{Fe}{2} $\lambda 4555$, Na~D, and
\ion{O}{1} $\lambda 7774$ lines have a velocity difference from their
smallest to largest wavelength component relative to the $gf$-weighted
wavelength for the feature of 2640, 8500, 300, and 130~km~s$^{-1}$,
respectively.
The Na~D and \ion{O}{1} $\lambda 7774$ lines should have the least
systematic errors in their measurements.  The velocity structure of
SN~2008ha indicates that the Fe is well distributed throughout the
ejecta and is not isolated in clumps.  In our $+62$~day spectrum, the
[\ion{Ca}{2}] lines are blueshifted by \about 400~km~s$^{-1}$ relative
to the rest frame and have full width at half-maximum intensity (FWHM)
of \about 1200~km~s$^{-1}$.

We find no evidence for \ion{He}{1} in any of our spectra; see
\S~\ref{ss:synow} for more details.

\subsection{Comparison of SN~2008ha to Other SNe}\label{ss:comp}

To better understand the nature of SN~2008ha, we have compared our
spectra of SN~2008ha to those of other peculiar SNe that show some
similarities to SN~2008ha.  We focus on SNe~1989B (a normal SN~Ia;
\citealt{Barbon90, Wells94}), 1991T (the prototype of its subclass of
SNe~Ia, and similar to SN~2002cx-like objects at early times;
\citealt{Filippenko92:91T, Phillips92}), 2002cx (the prototype of its
subclass of SNe; \citealt{Li03:02cx}), 2004aw (a peculiar SN~Ic that
was originally classified as a SN~Ia; \citealt{Taubenberger06}), 2005E
(a ``Ca-rich'' SN~Ib similar to prototypes SNe~2001co and 2003H;
\citealt{Filippenko03:carich, Foley05:05e}; that was both
underluminous and discovered in the halo of an S0/a galaxy; Perets
et~al., in prep.), 2005cs (a peculiar and low-luminosity SN~II;
\citealt{Pastorello06}), and 2007J (a SN which is very similar to
SN~2002cx at early times, but developed \ion{He}{1} lines at later
times; see the Appendix).

In Figure~\ref{f:comp}, we present our $+7$~day Lick spectrum of
SN~2008ha compared to the spectra of SNe~1991T and 2002cx.  We have
applied an inverse-variance weighted Gaussian smoothing function with
characteristic velocities of 800~km~s$^{-1}$ and 2000~km~s$^{-1}$
\citep{Blondin06} to our SN~2008ha spectrum.  To match the spectral
features of SN~2008ha, in addition to removing the recession
velocities of the SNe we remove an extra 7500 and 3000~km~s$^{-1}$ for
SNe~1991T and 2002cx, respectively.  This difference in characteristic
velocities correspond to a different kinetic energy per unit mass;
therefore, SN~2008ha has a lower kinetic energy per unit mass than
SN~2002cx, which has a lower kinetic energy per unit mass than
SN~1991T.

We see that the SN~2008ha spectrum smoothed with a 800~km~s$^{-1}$
Gaussian is very similar to SN~2002cx, indicating that SN~2008ha is
spectroscopically a member of the SN~2002cx subclass of SNe.  The
SN~2008ha spectrum convolved with a 2000~km~s$^{-1}$ Gaussian is also
relatively similar to the SN~1991T spectrum, but with SN~2008ha having
a redder continuum than SN~1991T.  The overall similarities between
these objects is the result of their optical spectra being dominated
by \ion{Fe}{2} lines at this phase.

\begin{figure}
\begin{center}
\epsscale{0.88}
\rotatebox{90}{
\plotone{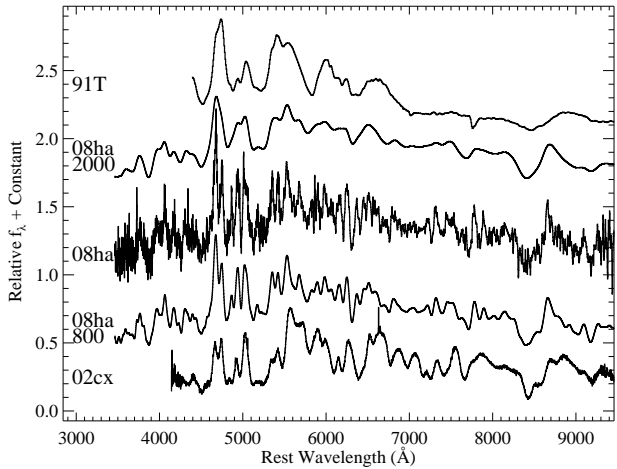}}
\caption{Optical spectra of SNe~1991T, 2002cx, and 2008ha.  All three
SN~2008ha spectra are the $+7$~day Lick spectrum, but two versions
have been smoothed with 800 and 2000~km~s$^{-1}$ Gaussians.  The
spectra of SNe~1991T and 2002cx both have phases of +20~days relative
to $B$ maximum.  The spectra of SNe~1991T and 2002cx have been
redshifted (after being deredshifted by their recession velocity) by
velocities of 7500 and 3000~km~s$^{-1}$, respectively.}\label{f:comp}
\end{center}
\end{figure}

\begin{figure}
\begin{center}
\epsscale{1.75}
\rotatebox{90}{
\plotone{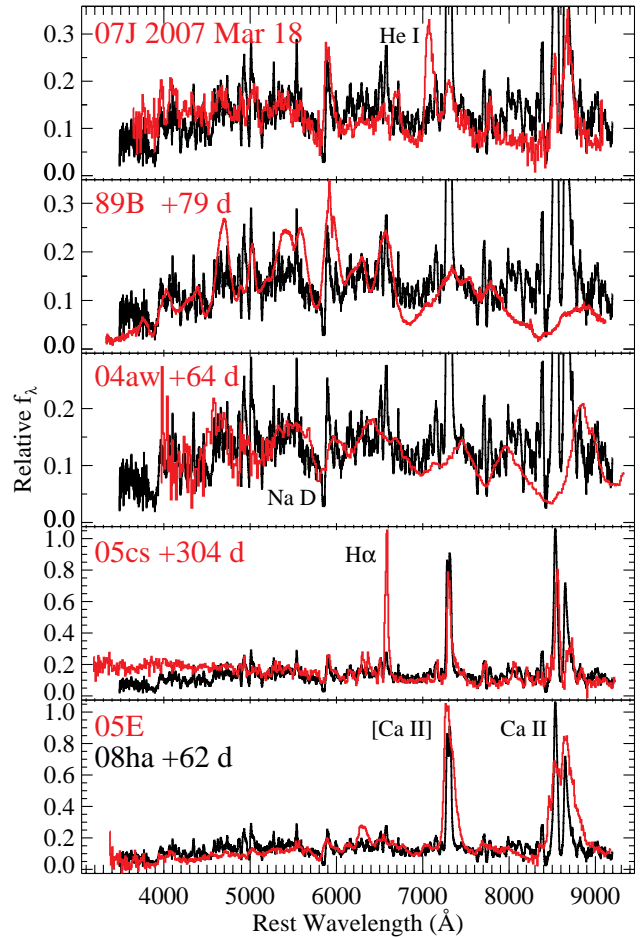}}
\caption{Late-time optical spectra of SNe~1989B, 2004aw, 2005E,
2005cs, 2007J, and 2008ha (black line in all panels).  The spectra of
SNe~1989B and 2004aw have been redshifted (after correcting for their
recession velocities) by a velocity of 3000 km~s$^{-1}$.  All spectra
have many of the same features, particularly \ion{Fe}{2} lines.  The
main differences are low line velocities for SNe~2005cs, 2007J, and
2008ha, H$\alpha$ emission for SN~2005cs, \ion{He}{1} emission for
SN~2007J, strong \ion{Ca}{2} and [\ion{Ca}{2}] emission for SNe~2005E,
2005cs, 2007J, and 2008ha, and [\ion{O}{1}] lines for SNe~2005E and
2005cs.  For clarity, we have interpolated over nebular emission lines
from the host galaxies.}\label{f:comp_late}
\end{center}
\end{figure}

In Figure~\ref{f:comp_late}, we present our $+62$~day spectrum of
SN~2008ha with comparison spectra of SNe~1989B, 2004aw, 2005E, 2005cs,
and 2007J at similar phases (with the exception of SN~2005cs).  Our
spectrum of SN~2005cs was obtained 304~days after explosion
\citep{Pastorello09}.  We have redshifted the spectra of SNe~1989B
and 2004aw (after removing their recession velocity) by a velocity of
3000 km~s$^{-1}$.  After this correction, we see that all six objects
are broadly consistent, having many similar spectral features.  In
particular, all objects have strong Na D and \ion{Fe}{2} lines.  The
main differences are that SN~2008ha has narrower lines than those of
the other objects, SNe~2005E, 2005cs, and 2008ha have strong
[\ion{Ca}{2}] and \ion{Ca}{2} emission lines, SNe~2004aw, 2005E and
2005cs have [\ion{O}{1}] emission, SN~2005cs has a strong H$\alpha$
line, and SN~2007J has prominent \ion{He}{1} emission lines.

The strong [\ion{Ca}{2}] and \ion{Ca}{2} lines in the latest SN~2008ha
spectrum present an interesting clue to the composition of the SN
ejecta.  The late-time ejecta of SNe~II cool through line emission
from H transitions.  Similarly, SNe~Ia and Ib/c cool through
[\ion{Fe}{2}] and [\ion{O}{1}] lines, respectively.  In all cases,
these lines dominate the late-time spectra of the objects.  Late-time
spectra of SN~2002cx show no strong forbidden emission lines (although
there are weak [\ion{Ca}{2}] lines and lines from other
intermediate-mass elements), indicating that the SN ejecta were still
quite dense at least a year after maximum light \citep{Jha06:02cx,
Sahu08}.  From our measured ejecta velocity and ejecta mass
measurement in \S~\ref{s:energy}, our $+62$~day spectrum should have
an electron density of \about $10^{9}$~cm$^{-3}$.  Integrating the
flux in the Ca lines, [\ion{Ca}{2}]/\ion{Ca}{2}~$ \approx 0.5$.  An
electron temperature of \about 7000~K is appropriate for this density
and line ratio \citep{Ferland89}.  At \about 200~days after maximum,
the electron density should be \about $10^{8}$~cm$^{-3}$, at which
point we should be able to measure the relative abundance of O/Ca
\citep{Fransson89}.  However, if the ejecta are in dense clumps, the
density may stay high for a longer time.

\subsection{SYNOW Model Fits}\label{ss:synow}

To investigate the details of our SN spectra, we use the supernova
spectrum-synthesis code SYNOW \citep{Fisher97}.  Although SYNOW has a
simple, parametric approach to creating synthetic spectra, it is a
powerful tool to aid line identifications which in turn provide
insights into the spectral formation of the objects. To generate a
synthetic spectrum, one inputs a blackbody temperature ($T_{\rm BB}$),
a photospheric velocity ($v_{ph}$), and for each involved ion, an
optical depth at a reference line, excitation temperature ($T_{\rm
exc}$), the maximum velocity of the opacity distribution ($v_{\rm
max}$), and a velocity scale ($v_{e}$). It assumes that the optical
depth declines exponentially for velocities above $v_{\rm ph}$ with an
$e$-folding scale of $v_{e}$.  The strengths of the lines for each ion
are determined by oscillator strengths and the approximation of a
Boltzmann distribution of the lower level populations with a
temperature of $T_{\rm exc}$.

\begin{figure*}
\begin{center}
\epsscale{0.88}
\rotatebox{270}{
\plotone{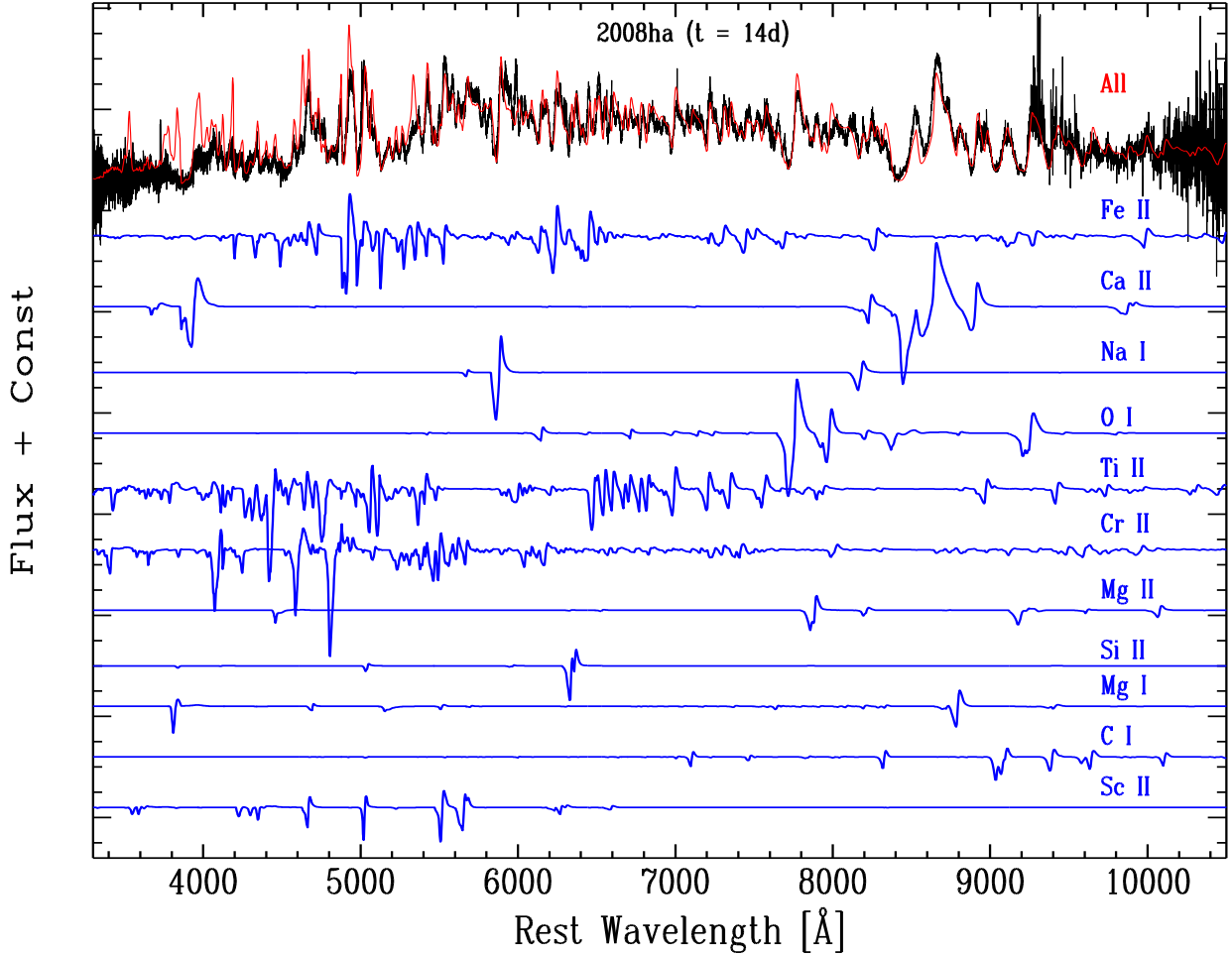}}
\caption{Our $+14$~day optical spectrum of SN~2008ha (black) and our
best-fit SYNOW synthetic spectrum (red).  In blue, we plot the
components (besides the blackbody continuum) of each synthetic
spectrum from each ion.  The components are plotted from most secure
identification at the top to least secure identification at the
bottom.  The 11-ion SYNOW spectrum reproduces almost all features in
the spectrum of SN~2008ha.}\label{f:synow_08ha}
\end{center}
\end{figure*}

In Figure~\ref{f:synow_08ha}, we present our $+14$~day spectrum of
SN~2008ha with a synthetic spectrum generated from SYNOW.  This fit
has $T_{\rm BB} = 5500$~K and $v_{\rm ph} = 600$~km~s$^{-1}$.  Eleven
ions are used in this fit. Several ions, which we consider as
``secure'' identifications, are important for the spectral formation
of SN~2008ha.  These ions account for either the most prominent or
multiple line features. The secure ions include \ion{Fe}{2},
\ion{Ca}{2}, \ion{Na}{1}, \ion{O}{1}, \ion{Ti}{2}, and \ion{Cr}{2}. An
additional 5 ions (\ion{Mg}{2}, \ion{Si}{2}, \ion{Mg}{1}, \ion{C}{1},
and \ion{Sc}{2}) are considered to be ``possible" identifications, as
each of them fit one line or multiple weak lines.  Because SN~2008ha
has exceptionally low expansion velocities and narrow line widths,
many otherwise blended lines are resolved. The fact that our SYNOW fit
provides an excellent fit to over 70 line features suggests that we
have identified the most important ions that contribute to the
spectral formation of SN~2008ha.

\citetalias{Valenti09} presented a SYNOW fit to a spectrum taken a day
after our $+14$~day spectrum.  Their fit used the parameters $T_{\rm
BB} = 6200$~K and $v_{\rm ph} = 1300$~km~s$^{-1}$.  Both of these
values are higher than our fit.  They also include \ion{He}{1} while
excluding \ion{Mg}{1} and \ion{C}{1}.  We find that inclusion of
\ion{Mg}{1} and \ion{C}{1} improves the fit more than \ion{Sc}{2};
thus, \ion{Mg}{1} and \ion{C}{1} are more secure.  We have attempted a
fit that includes \ion{He}{1}, and while it improves the fit in some
wavelength regions, it makes it worse in others.  Overall, we believe
that adding \ion{He}{1} to our particular model spectrum made it
appear less like the spectrum of SN~2008ha.

If we simply take the ions used in our SN~2008ha model and increase
the photospheric velocity to match that of SN~2002cx, the SYNOW fit is
not a good match to SN~2002cx.  However, if we take the SN~2008ha
model, remove \ion{C}{1}, reduce the strength of \ion{Ca}{2}, and
increase the strength of \ion{Fe}{2}, \ion{Co}{1}, and \ion{Co}{2}, we
can achieve a good match to SN~2002cx.  This indicates that SN~2008ha
has lower expansion velocities at the photosphere, stronger
intermediate-mass element lines, and weaker Fe-peak element lines.
These differences can all be explained by a lower energy explosion
with less $^{56}$Ni production in SN~2008ha.

\begin{figure}
\begin{center}
\epsscale{0.88}
\rotatebox{270}{
\plotone{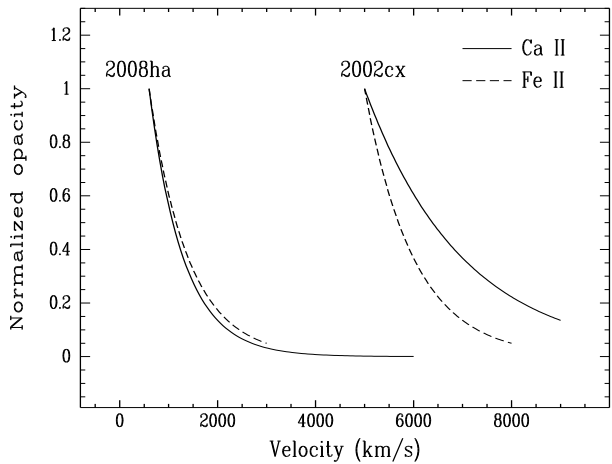}}
\caption{Normalized opacity distribution in velocity space for
SNe~2002cx and 2008ha as determined from SYNOW fits. \ion{Ca}{2}
(solid lines) represents intermediate-mass elements, while \ion{Fe}{2}
(dashed lines) represent Fe-peak elements.}\label{f:synow_op}
\end{center}
\end{figure}

Because of the highly parameterized nature of the SYNOW fits, the
derived quantities of the input parameters are often not unique and
the associated uncertainties are difficult to estimate.  Nonetheless,
the distribution of the opacities in velocity space can provide useful
information on the spectral formation of the SNe.  In
Figure~\ref{f:synow_op}, we plot the opacity distribution of two
representative ions, \ion{Ca}{2} for the intermediate-mass elements
and \ion{Fe}{2} for the Fe-peak elements. The opacities are normalized
to be unity at the photosphere (to facilitate the making of the plot)
and the high-velocity end of the curves marks $v_{\rm max}$. While the
\ion{Ca}{2} opacity shows a considerably flatter distribution than
\ion{Fe}{2} in SN~2002cx, the two distributions are nearly
identical in SN~2008ha.  The \ion{Ca}{2} opacity distribution shows
some overlap in velocity space among the SNe (SN~2008ha has
significant opacity in the range 4000--6000~km~s$^{-1}$ because it has
an enormous opacity at the photosphere).  The \ion{Fe}{2} opacity
distribution, on the other hand, is nearly ``stratified"; the opacity
is only present between 600 and 3000~km~s$^{-1}$ for SN~2008ha and
5000 to 8000~km~s$^{-1}$ for SN~2002cx.  The differences in the
opacity distributions may offer clues to the nature of the explosions.

\section{Energetics}\label{s:energy}

\begin{figure}
\begin{center}
\epsscale{1.3}
\rotatebox{90}{
\plotone{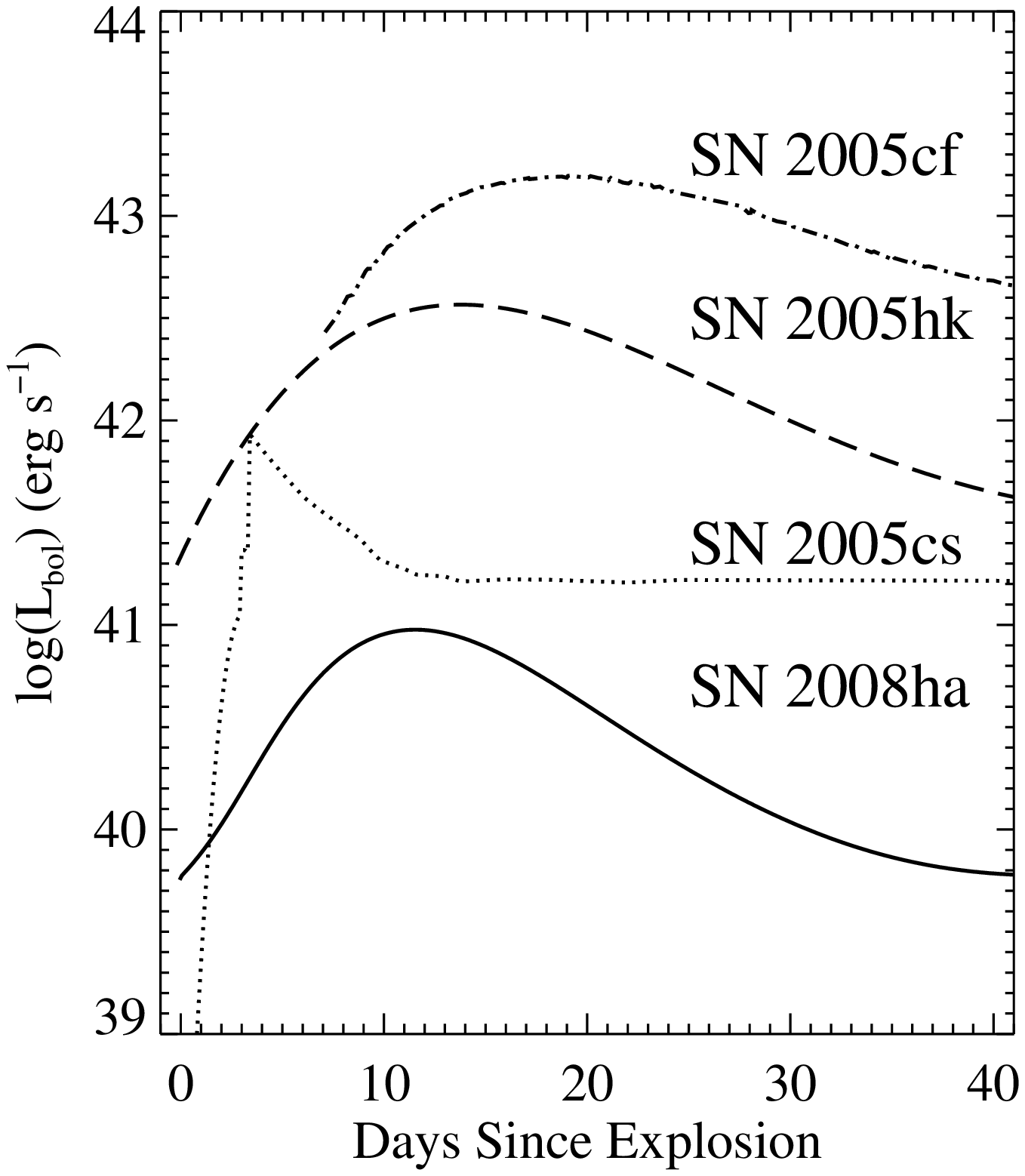}}
\caption{Quasi-bolometric light curve of SNe~2005cf (dot-dashed line),
2005cs (dotted line), 2005hk (dashed line), and 2008ha (solid
line).}\label{f:bolo}
\end{center}
\end{figure}

Using our \bvrijhk light curves, we are able to construct a
quasi-bolometric light curve for SN~2008ha covering the wavelength
range $3500 \le \lambda < \infty$.  We fit polynomials to our light
curves to interpolate for dates where we have no photometry.  We fit a
blackbody spectrum to the \vrijhk data and extrapolate this fit for
wavelengths longer than the $K$ band.  Our resulting light curve is
presented in Figure~\ref{f:bolo}.  We perform the same procedure for
SN~2005hk (except without the $K$ band), using the light curves from
\citet{Phillips07}.  The SN~2005hk bolometric light curve from
\citet{Phillips07} is slightly different than the one we present.  We
used \bvrijh bands, while they used \ugrizyjh and included unpublished
{\it Swift} UV photometry.  To compensate for the lack of UV
information, we have scaled the $B$-band flux so that the peaks of our
light curve and that of \citet{Phillips07} have approximately the same
luminosity at peak.  For consistency, the same correction is applied
to SN~2008ha.

The same technique is applied to the light curves of the underluminous
SN~II 2005cs \citep{Pastorello09} with the one difference that for
regions where the light curve has good coverage, the light curves are
interpolated between points rather than being fit by a polynomial,
which does not produce a good fit to the peak and plateau stages
simultaneously.  The bolometric light curve of the normal SN~Ia 2005cf
\citep{Wang09} is also included.

Our bolometric light curve of SN~2008ha is poorly constrained before
maximum brightness, where our photometric coverage is sparse.
Similarly, beyond \about 40~days after explosion (\about 30~days after
$B$ maximum) the light curve is poorly constrained.  Additional
late-time data will improve the light curve by constraining the
late-time decay rate.

From Figure~\ref{f:bolo}, we see that SN~2008ha is an extremely
low-luminosity event, with a peak luminosity of $9.5 \pm 1.4 \times
10^{40}$~erg~s$^{-1}$.  This indicates that at peak, SNe~2005cf and
2005hk were 170 and 40 times more luminous than SN~2008ha,
respectively.  Ignoring the SN~IIn imposters, which are likely
luminous outbursts of luminous blue variables \citep[LBVs;
e.g.,]{VanDyk03, Maund06, Berger09, Smith09}, SN~2008ha is perhaps the
least luminous SN ever observed (SN~1999br had a \bvri
quasi-bolometric peak luminosity of $5 \times 10^{40}$, but a
significant percentage of its luminosity may have been emitted at UV
or NIR wavelengths; \citealt{Pastorello04}).  Integrating the light
curves, we see that over the first 30~days after explosion, SN~2008ha
emitted $E_{\rm rad} = 1.2 \times 10^{47}$~ergs, while SN~2005hk
emitted 50 times more energy ($E_{\rm rad} = 5.7 \times 10^{48}$~ergs)
and SN~2005cs emitted 5 times more energy ($E_{\rm rad} = 5.5 \times
10^{47}$~ergs).

While the bolometric light curves of SNe~2005hk and 2008ha are
declining at 30~days after explosion, the bolometric light curve of
SN~2005cs is relatively flat at 30~days after explosion, maintaining
that luminosity until \about 100~days after explosion.  The late-time
energy emission can contribute significantly to the overall energy
emitted by SN~2005cs, while the bulk of the energy emitted by
SNe~2008ha and 2005hk is well represented by the early-time bolometric
light curves.  SN~2005cs emitted a similar amount of energy as
SN~2008ha over the first month after explosion; however, the peak
luminosity of SN~2008ha is smaller than the luminosity of SN~2005cs on
its plateau.

Making some simple assumptions about the SN ejecta and assuming that
the light curve is predominantly powered by $^{56}$Ni, we can
determine the $^{56}$Ni mass from the rise time and peak of the
bolometric light curve \citep{Arnett82}.  Using a rise time of 10~days
and a peak luminosity of $9.5 \times 10^{40}$~ergs~s$^{-1}$, we derive
a $^{56}$Ni mass of $(3.0 \pm 0.9) \times 10^{-3}$ M$_{\sun}$ (with
the majority of the error coming from the uncertainty in the rise
time).  However, the decay of $^{56}$Ni may not be the main source of
energy.  Alternatively, the SN may be powered by other radioactive
nuclei or circumstellar interaction (although no evidence of this is
observed).

We can break the degeneracy of $M_{\rm ej}$ and $E_{\rm kin}$ by
examining the rise time of the light curves.  While the ejecta
velocity is proportional to $(E_{\rm kin} / M_{\rm ej})^{1/2}$, the
rise time of a SN light curve is proportional to $(M_{\rm ej}^{3} /
E_{\rm kin})^{1/4}$ \citep{Arnett82}.  Combining these equations and
assuming that two objects have the same opacity, we have
\begin{equation}
  E_{{\rm kin, } 1} / E_{{\rm kin, } 2} = \left ( \frac{v_{1}}{v_{2}}
    \right )^{3} \left ( \frac{t_{1}}{t_{2}} \right )^{2}
\end{equation}
and
\begin{equation}
  M_{{\rm ej}, 1} / M_{{\rm ej, } 2} = \frac{v_{1}}{v_{2}} \left (
    \frac{t_{1}}{t_{2}} \right )^{2}.
\end{equation}
\noindent
Using a SN~Ia as a reference with $t_{r} = 19.5$~days and $v =
10,000$~km~s$^{-1}$, we find $E_{\rm kin, 08ha} / E_{\rm kin, Ia} = 2.3
\times 10^{-3}$ and $M_{\rm ej, 08ha} / M_{\rm ej, Ia} = 0.11$.
Assuming $E_{\rm kin, Ia} = 10^{51}$~ergs and $M_{\rm ej, Ia} = 1.4$
M$_{\sun}$, we find $E_{\rm kin, 08ha} = 2.3 \times 10^{48}$~ergs and
$M_{\rm ej, 08ha} = 0.15$ M$_{\sun}$.

Additionally, we can measure the ejecta mass directly if we know the
opacity.  From \citet{Arnett82} and \citet{Pinto00}, we find
\begin{equation}\label{e:ej}
  M_{\rm ej} = 0.16 \left ( \frac{t_{r}}{10~{\rm d}} \right )^{2} \left (
    \frac{0.1~ {\rm cm}^{2}{\rm g}^{-1}}{\kappa} \right ) \left (
    \frac{v}{2 \times 10^{8}~ {\rm cm~s}^{-1}} \right ),
\end{equation}
where $t_{r}$ is the rise time and $\kappa$ is the opacity.  A value
of $\kappa = 0.1$~cm$^{2}$~g$^{-1}$ is reasonable for a spectrum
dominated by \ion{Fe}{2} lines \citep{Pinto00}.  This value is
consistent with what we found above; however, we gained no new
information, since both assumed $\kappa \approx
0.1$~cm$^{2}$~g$^{-1}$.  A smaller value of $\kappa$ would result in a
larger ejecta mass.

As discussed in \S~\ref{s:phot}, \citetalias{Valenti09} derived a peak
absolute $R$-band magnitude that is 0.2~mag brighter than what we have
measured.  They then scaled the $^{56}$Ni mass derived for SN~2005hk
\citep{Phillips07} by the difference in the luminosities of SNe~2005hk
and 2008ha in the $R$ band to derive $M_{^{56}{\rm Ni}} = 5 \times
10^{-3}$.  Although they ignored the difference in rise times between
SNe~2005hk and 2008ha, our derived values for $M_{^{56}{\rm Ni}}$ are
similar because the rise-time effect is partially canceled by the
difference in luminosity.  They also derived a kinetic energy of
$E_{\rm kin} = 5 \times 10^{49}$~ergs, an order of magnitude larger
than what we have determined, and an ejecta mass of $M_{\rm ej} =
0.2$--0.7 M$_{\sun}$, slightly larger than what we have found.  They
do not measure a rise time for SN~2008ha, so presumably for their
derived values they used the rise time of SN~2005hk.  They also used
their measured value of $v = 2300$~km~s$^{-1}$, which is slightly
larger than our value of 2000~km~s$^{-1}$.  Using their values, we
would measure a kinetic energy and ejecta mass 3.1 and 2.3 times
larger than our reported values, respectively.  Making these changes
brings their values closer to ours, but the kinetic energy is still
discrepant.


\section{Host Galaxies}\label{s:hosts}

Examining the host galaxies of SNe can indicate realistic potential
progenitors.  For instance, SNe~Ia are routinely found in elliptical
galaxies, indicating that at least some of them come from an older
stellar population.  \citetalias{Valenti09} examined the host-galaxy
morphology distribution of several SN~2002cx-like objects, finding
that all 8 SNe in their sample (including SN~2007J, which may not be a
true member of the class; see the Appendix) were in late-type hosts
(Sb or later).  They then claim that the probability that SNe~Ia and
SN~2002cx-like objects share the same parent population is between 5\%
and 14\%.  However, they did not include SNe~2008A
\citep{Blondin08:08A}, 2008ae \citep{Blondin08:08ae}, 2008ge
\citep{Stritzinger08}, or 2009J \citep{Stritzinger09}, all of which
have been classified as SN~2002cx-like objects in the literature.  Two
of the four objects that they did not include (SNe~2008ae and 2008ge)
have S0 hosts, while another (SN~2008A) has an Sa host.

In this Section, we will first add three previously misclassified
members to the SN~2002cx class.  With this extended sample, we will
compare the host-galaxy morphology distribution of SN~2002cx-like
objects to other SN classes.

\subsection{The Extended Sample of SN~2002cx-like Objects}

\begin{figure}
\begin{center}
\epsscale{0.85}
\rotatebox{90}{
\plotone{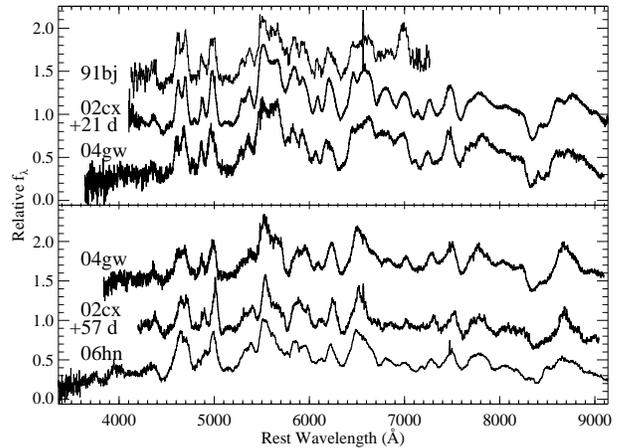}}
\caption{({\it top}): Optical spectra of SNe~1991bj, 2002cx (at a
phase of $t = +21$~days relative to $B$ maximum), and 2004gw.  ({\it
bottom}): Optical spectra of SNe~2002cx (at a phase of $t = +57$~days
relative to $B$ maximum), 2004gw, and 2006hn.  The spectra of
SNe~1991bj, 2004gw, and 2006hn have all be dereddened by a
\citet{Cardelli89} extinction law with $E(B-V) = 0.25$, 0.40, and
0.30~mag, respectively, and $R_{V} = 3.1$.  The spectra of SN~2004gw
are separated by 28~days, while the spectra of SN~2002cx are separated
by 36~days.  SNe~1991bj, 2004gw, and 2006hn are all members of the
SN~2002cx class of objects.}\label{f:91bj-06hn}
\end{center}
\end{figure}

In the literature, there are currently 10 SNe identified as
SN~2002cx-like objects.  In Figure~\ref{f:91bj-06hn}, we present
optical spectra of SNe~1991bj, 2004gw, and 2006hn, all of which we
propose are new members of this class.

The spectra of SNe~1991bj and 2006hn were obtained with the UV Schmidt
and Kast spectrographs, respectively, mounted on the Lick 3~m Shane
telescope.  The spectra of SN~2004gw were obtained with LRIS mounted
on the Keck~I telescope.  Classifications based on these spectra were
originally reported by \citet{Pollas92}, \citet{Foley05:05e} and
\citet{Filippenko05:04gw}, and \citet{Foley06:06hn} for SNe~1991bj,
2004gw, and 2005hn, respectively.  SN~1991bj was recently identified
as a potential member of this class \citep{Stanishev07}.  SN~2004gw
was originally classified as a probable SN~Ia with ``a number of
spectral peculiarities'' \citep{Foley05:05e}, but \citet{Gal-Yam05}
suggested that it may be a SN~Ic.  Finally, \citet{Filippenko05:04gw}
reclassified the object as a SN~Ia.  Additional observing details can
be found in those references as well as in \citet{Foley09:lowz}.  In
Figure~\ref{f:91bj-06hn}, we compare the spectra of these objects to
spectra of SN~2002cx.  The spectra shown here demonstrate clearly that
each of these SNe belongs to the SN~2002cx class.  (Incidentally, a
striking number of peculiar SN classes had their first members
discovered in 1991.)  In addition to these 14 objects, there is one
potential member: SN~2007J.

SN~2007J, which had many similarities to SN~2002cx at early times
\citep{Filippenko07:07J1}, eventually developed He emission lines
\citep{Filippenko07:07J2}, unlike any other SN in this class.  We
discuss SN~2007J in detail in the Appendix.  Although this object has
many similarities to SN~2002cx, the presence of \ion{He}{1} lines in
its spectra casts doubt on its inclusion in this class.  SN~2008ha
also has some significant differences from SN~2002cx.  However, we
believe that these differences are caused mainly by the low-energy
explosion in SN~2008ha rather than differences in progenitors.

\subsection{Host Galaxies of SN~2002cx-like Objects}\label{s:02cx}

Using the extended sample of SN~2002cx-like objects, we investigate
the host galaxies of the this class.  In Table~\ref{t:host}, the
detailed host information for 14 members of this class and 1 potential
member is presented.  We exclude SN~2007J from the host-galaxy
analysis, but its inclusion does not significantly change the results.

\begin{deluxetable*}{lcll}
\tabletypesize{\scriptsize}
\tablewidth{0pt}
\tablecaption{Host-Galaxy Properties of SN~2002cx-like Objects\label{t:host}}
\tablehead{
\colhead{SN Name} &
\colhead{Reference} &
\colhead{Host-Galaxy Name} &
\colhead{Morphology}}

\startdata

1991bj                   & 1,2,3    & IC~344                   & Sb  \\
2002cx                   & 4,5,6    & CGCG~044-035             & Sb  \\
2003gq                   & 7,8      & NGC~7407                 & Sbc \\
2004gw                   & 1,9,10   & PGC 16812                & Sbc \\
2005P                    & 6        & NGC~5468                 & Scd \\
2005cc                   & 11       & NGC~5383                 & Sb  \\
2005hk                   & 12,13,14 & UGC~272                  & Sd  \\
2006hn                   & 1,15     & NGC~6154                 & Sa  \\
2007J\tablenotemark{a}   & 1,16,17  & UGC~1778                 & Sd  \\
2007qd                   & 18       & SDSS~J020932.74-005959.6 & Sc  \\
2008A                    & 19       & NGC~634                  & Sa  \\
2008ae                   & 20       & IC~577                   & S0  \\
2008ge                   & 21       & NGC~1527                 & S0  \\
2008ha                   & 1,22,23  & UGC~12682                & Irr \\
2009J                    & 24       & IC~2160                  & Sbc

\enddata

\tablerefs{1 = This paper, 2 = \citet{Pollas92}, 3 =
\citet{Stanishev07}, 4 = \citet{Li03:02cx}, 5 = \citet{Branch04}, 6 =
\citet{Jha06:02cx}, 7 = \citet{Filippenko03:03gq1}, 8 =
\citet{Filippenko03:03gq2}, 9 = \citet{Foley05:05e}, 10 =
\citet{Filippenko05:04gw}, 11 = \citet{Antilogus05}, 12 =
\citet{Chornock06}, 13 = \citet{Phillips07}, 14 = \citet{Sahu08}, 15 =
\citet{Foley06:06hn}, 16 = \citet{Filippenko07:07J1}, 17 =
\citet{Filippenko07:07J2}, 18 = \citet{Goobar07}, 19 =
\citet{Blondin08:08A}, 20 = \citet{Blondin08:08ae}, 21 =
\citet{Stritzinger08}, 22 = \citet{Foley08:08ha}, 23 =
\citetalias{Valenti09}, 24 = \citet{Stritzinger09}.}

\tablenotetext{a}{Shows \ion{He}{1} emission lines at late times and
may not be a true member of the class.  It has been removed from the
sample when discussing host-galaxy properties.}

\end{deluxetable*}

None of the SN~2002cx-like objects have elliptical hosts.  With six
SN~2002cx-like objects and two potential members,
\citetalias{Valenti09} made the claim that all SN~2002cx-like objects
have late-type hosts; however, with our extended sample, this claim
does not hold.  In our extended sample, which has twice the number of
definitive SN~2002cx-like objects than the sample presented by
\citetalias{Valenti09}, there is 1/14 (3/14) objects with S0 (S0
through Sa) host galaxies.

\begin{figure}
\begin{center}
\epsscale{1.2}
\rotatebox{90}{
\plotone{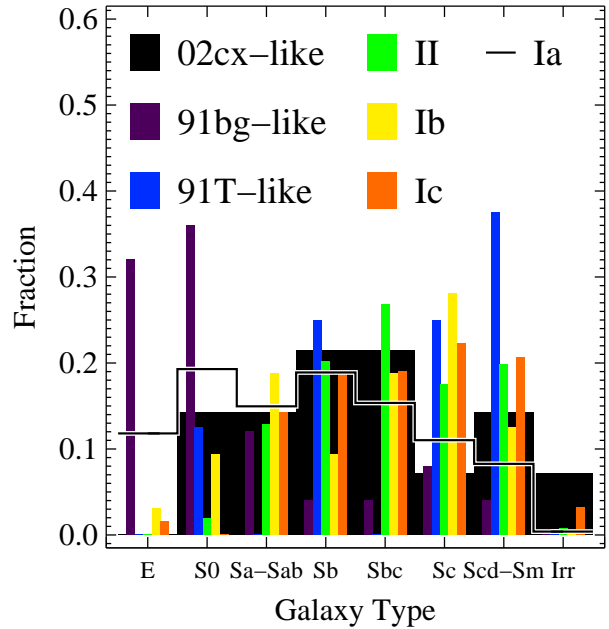}}
\caption{Fraction of host galaxies in a given morphological bin
for the (sub)classes of SN~2002cx-like objects (black histogram),
SN~1991bg-like objects (purple histogram), SN~1991T-like objects (blue
histogram), SNe~II (green histogram), SNe~Ib (yellow histogram),
SNe~Ic (orange histogram), and normal SNe~Ia (black
line).}\label{f:frac}
\end{center}
\end{figure}

A comparison sample was generated using the SNe found in galaxies
monitored by KAIT (which includes the host of SN~2008ha).  Comparing
the host-galaxy morphology of various (sub)classes of SNe may indicate
similarities in progenitors.  For consistency, we use the host-galaxy
classification scheme presented in \citet{Leaman09}, derived mainly
from NED, and we divide the galaxies into eight morphology bins.  In
Figure~\ref{f:frac}, the fraction of host galaxies in these eight
morphological bins for several SN classes is presented.

The well-known results that SN~1991bg-like objects
\citep{Filippenko92:91bg, Leibundgut93} tend to be found in early-type
galaxies (relative to normal SNe~Ia; \citealt{Howell01}), while
SNe~Ib, Ic, II, and 1991T-like SNe have distributions skewed to
later-type host galaxies relative to that of normal SNe~Ia are
immediately apparent in Figure~\ref{f:frac}.  SN~2002cx-like events
have, on average, later-type host galaxies than normal SNe~Ia.

Examining Figure~\ref{f:frac}, all distributions other than that of
SN~1991bg-like SN host galaxies are somewhat similar.  Performing a
Kolmogorov-Smirnov (K-S) test on the data, we find that SN~1991bg-like
objects, SNe~II, and SNe~Ic have significantly different distributions
from normal SNe~Ia.  SNe~Ib and SN~1991T-like objects have marginally
different distributions from normal SNe~Ia.  In a similar study,
\citet{Kelly08} found that the locations of SNe~Ia, Ib, and II all
followed the galaxy light, while SNe~Ic were more likely to be found
in the brightest regions of a galaxy, indicating more massive
progenitors for SNe~Ic than SNe~Ib or II.  Except for SN~1991bg-like
objects, no SN class has a significantly different host-galaxy
morphology distribution from SN~2002cx-like objects.  Although these
results may change with a slightly different sample (KAIT is biased to
luminous galaxies) or with increased numbers, we cannot say that
SN~2002cx-like objects have a different host-galaxy morphology
distribution from any SN class other than SN~1991bg-like events.

Using the sample presented by \citetalias{Valenti09}, we find that
SN~2002cx-like objects have a distribution marginally inconsistent
with that of normal SNe~Ia.  We also find that their host-galaxy
morphology distribution is very similar to that of the SN~1991T-like
objects from the KAIT sample.  Therefore, even with the incomplete
sample of \citetalias{Valenti09}, SN~2002cx-like objects have a consistent
host-galaxy morphology distribution to one subclass of SNe~Ia.

We find that 12\% of the normal SN~Ia hosts in the KAIT sample are
elliptical galaxies.  If SN~2002cx-like events have the same
host-morphology distribution as normal SNe~Ia, we would expect to have
found $1.7 \pm 1.3$ SNe ($1.8 \pm 1.3$ if we include SN~2007J) of the
SN~2002cx class in elliptical galaxies, consistent with our finding
zero members in elliptical galaxies.  We find that 31\% of SNe~Ia come
from E through S0/a galaxies.  If we assume that SN~2002cx-like
objects and normal SNe~Ia have the same host-galaxy morphology
distribution, then we would expect $4.4 \pm 2.1$ SNe ($4.7 \pm 2.2$ if
we include SN~2007J) of the SN~2002cx class in early-type galaxies.
This prediction is consistent with our measurement of 2 members in
early-type galaxies.

\subsection{The Host Galaxy of SN~2008ha}

In addition to the morphology measurements for all SN~2002cx-like host
galaxies, we have made additional measurements of UGC~12682, the host
of SN~2008ha.

\citet{Takamiya95} measured UGC~12682 to have $V = 13.84$~mag and
$B-V = 0.28$~mag within an aperture of 84.5\arcsec\ radius.  With our
adopted distance modulus and correcting for Milky Way extinction
\citep{Schlegel98}, we find $M_{V} = -18.0$~mag, comparable to the
absolute magnitude of the Large Magellanic Cloud (LMC; $M_{V} = 
-18.4$~mag).

For our LRIS observation, we positioned the slit to go through a
bright knot in UGC~12682.  The emission from the host galaxy in a
16$\arcsec$ aperture around this bright knot displays intensity ratios
of [\ion{N}{2}] $\lambda 6584$/H$\alpha = 0.057$ and [\ion{O}{3}]
$\lambda 5007$/H$\beta = 4.0$.  The N2 and O3N2 metallicity indicators
of \citet{Pettini04} give consistent estimates for the host-galaxy
oxygen abundance of $12 + \log({\rm O/H}) = 8.16 \pm 0.15$, well below
the solar value of 8.9 or 8.7 \citep{Delahaye06, Asplund05}.

We can use the far-infrared luminosity of UGC~12682 to estimate its
star formation rate (SFR).  Assuming a distance of 21.3~Mpc and the
\emph{IRAS} measurements of the 60 and 100~$\mu$m flux
\citep{Melisse94}, the calibration of \citet{Kennicutt98} yields an
estimate for the SFR of 0.07 M$_{\sun}$~yr$^{-1}$.  Although this
seems low for a ``star forming'' galaxy, it is consistent with its
luminosity.  Using the same calibration for the far-IR luminosity,
\citet{Whitney08} find a SFR for the LMC of 0.17 M$_{\sun}$~yr$^{-1}$,
about 2.5 times larger than that of UGC~12682 (while the LMC is about
1.5 times more luminous).


\section{Model Comparisons}\label{s:model}

With our extensive data set, derived values for the energetics,
host-galaxy properties, and knowledge of the host-galaxy morphology
distribution of SN~2002cx-like objects, we can investigate various
scenarios which may lead to a SN similar to SN~2008ha.  These models
can generally be separated into two classes: core-collapse and
thermonuclear explosions.  In the former, the progenitor is a massive
star whose internal pressure support is surpassed by gravitational
pressure, causing a collapse and subsequent explosion.  The latter
class requires a thermonuclear explosion of degenerate matter.  We
certainly have not examined all possible models, but we have addressed
some of the more prevalent models for exotic SNe.

Based on their light curve, spectra, and host-galaxy information,
\citetalias{Valenti09} claim that SN~2008ha and all SN~2002cx-like objects
are core-collapse events.  Specifically, they suggest that SN~2008ha
was either the result of a fallback event or an electron-capture event
of the O-Ne-Mg core of a 7--8 M$_{\sun}$ star.  We will examine both
of these scenarios in addition to other models which may explain our
observations of SN~2008ha.

A correct model for SN~2008ha must reproduce the following
observations: a peak luminosity of \about $10^{41}$~erg~s$^{-1}$, a
rise time of \about 10~days, ejecta velocities of \about
2000~km~s$^{-1}$, an ejecta mass of 0.15 M$_{\sun}$, a $^{56}$Ni mass
(assuming that the light curve is powered by the radioactive decay of
$^{56}$Ni) of $4 \times 10^{-3}$ M$_{\sun}$, and a kinetic energy of
\about $2 \times 10^{48}$~ergs.  Additionally, if all SN~2002cx-like
objects including SNe~2007J and 2008ha are to be explained by one
model, it must explain the occurrence of some objects in S0 galaxies,
large ranges of luminosity, rise time, and energy, and the presence of
He in the progenitor star and/or circumstellar environment.  With
these constraints in mind, we examine each model in turn.

\subsection{Direct Collapse/Fallback}

It has been suggested that during the gravitational collapse of a
massive star, the entire star may collapse directly to a black hole
instead of creating a shock wave that disrupts the outer layers of the
star.  Alternatively, a massive star could produce a proto-neutron
star near the limiting mass during collapse, which would then quickly
create a black hole by accreting some material which falls back onto
the proto-neutron star.  These processes would not produce any or much
electromagnetic radiation, resulting in a faint (or potentially no)
SN.

The progenitors of these systems must be very massive stars to produce
a black hole.  Although other factors such as metallicity, binarity,
or rotation may determine if a particular star will directly collapse
to a black hole, most estimates suggest that an initial mass of
$\gtrsim 40$ M$_{\sun}$ is necessary \citep{Heger03}.  Similarly, an
initial mass of $\gtrsim 30$ M$_{\sun}$ is necessary to create a black
hole through fallback \citep{Fryer99}.

Massive SN~Ic progenitor stars lose their outer hydrogen and helium
layers through either stellar winds/outbursts or binary transfer.  For
a single-star system, the amount of mass loss through winds is
directly dependent on the metallicity of the star.  \citet{Modjaz08}
found that the sites of broad-lined SNe~Ic (a subclass of SNe~Ic with
expansion velocities up to \about $0.1 c$) that are not associated
with GRBs have metallicities similar to solar metallicity.  The
progenitors of weak SNe~Ic (where a black hole is formed through
fallback) should have lower metallicities than that of the progenitors
of normal SNe~Ic \citep{Heger03}.

Since the progenitors of black holes must be very massive stars, we
expect the distribution of their associated SNe to be heavily biased
to star-forming galaxies.  The host of SN~2008ha is an irregular
galaxy.  In \S~\ref{s:hosts}, we found for UGC~12682, the host of
SN~2008ha, a subsolar oxygen abundance of $12 + \log({\rm O/H}) =
8.16$ and a SFR of 0.07 M$_{\sun}$~yr$^{-1}$, which we determined was
large considering its luminosity is comparable to that of the LMC.  A
40 M$_{\sun}$ star with subsolar metallicity has a lifetime of \about
$5 \times 10^{6}$~yr \citep{Schaller92}.  With our measured SFR and an
LMC initial-mass function \citep{Massey95}, we calculate that there
are 120 stars with $M \gtrsim 40$ M$_{\sun}$ in UGC~12682 at any given
time.  Using the lifetime of these stars, we estimate a SN rate of $2
\times 10^{-5}$~yr$^{-1}$ for these stars.

All observations of SN~2008ha can be explained through this model if
we assume that it had the precise initial conditions necessary to
barely produce an explosion, rather than directly collapse to a black
hole.  \citetalias{Valenti09} make no specific claims regarding this
scenario.  For the class of SN~2002cx-like objects, this model would
explain the diversity of the class.  While most objects in this class
are hosted by a range of star-forming galaxies, one of these objects
have S0 hosts.  We do not expect there to be a large population of
massive stars in these galaxies (although a more thorough search for
current star formation in this particular host is necessary).  It is
therefore unlikely that this model is the mechanism for all members of
the SN~2002cx-like class.

\subsection{Electron Capture}

Stars with an initial mass of \about9~M$_{\sun}$ will evolve to a
super-AGB stage with a massive, degenerate, O-Ne-Mg core.  These stars
may evolve further to become O-Ne WDs.  Either in the AGB or white
dwarf (WD) phases, electron capture can then induce a core-collapse
event, triggering a SN \citep[e.g.,][]{Miyaji80, Nomoto84:ec,
Kitaura06, Metzger08}.  These models predict a small $^{56}$Ni mass
($10^{-2}$--$10^{-1}$ M$_{\sun}$) and explosion energy (\about
$10^{50}$ ergs) \citep{Kitaura06}.

Since there is no hydrogen present in the spectra of SN~2008ha, its
progenitor could not have been an AGB star.  Similarly, the
progenitors of SN~2002cx-like objects could not be AGB stars.
However, this intermediate-mass range for progenitor stars is
consistent with the host-galaxy morphology distribution of
SN~2002cx-like events (see \S~\ref{s:02cx}).  \citetalias{Valenti09}
suggest that electron capture of a star with a O-Ne-Mg core is a
possible model for SN~2008ha.  However, we conclude that if electron
capture is the correct model for SN~2008ha, the progenitor must have
been a WD.

\citet{Metzger08} presents a model for a single WD collapsing to a
neutron star through electron capture.  In this model, they predict
the creation of $10^{-2}$ M$_{\sun}$ of $^{56}$Ni ejected at
relativistic velocities, but essentially no intermediate-mass
elements.  However, if two WDs merge and subsequently collapse to a
neutron star, there will be material at the radius of the recently
formed WD and at the radius of the newly formed neutron star.  The
material around the neutron star would produce the same ejecta with
the same energy as the single-star model, but the explosion would
occur within the WD accretion disk, mass-loading the explosion,
reducing the kinetic energy per unit mass, increasing the opacity (and
thus the time scale of the event), and heating the WD accretion disk
to temperatures where intermediate-mass elements may form (but not hot
enough to produce additional $^{56}$Ni).

The conventional electron-capture scenario of an AGB star does not fit
the observations of any of the SN~2002cx-like objects.  If the
progenitor is a single WD, the model fails to produce the
intermediate-mass elements or low velocities seen for these objects.
However, if there is a WD merger followed by the accretion-induced
collapse of the newly formed WD, we may reproduce the observations of
all SN~2002cx-like objects.  Additionally, if one of the objects was a
He WD, this model may explain the \ion{He}{1} lines seen in the
spectra of SN~2007J.

\subsection{Deflagration}

It has been suggested that SN~2002cx-like objects are the result of a
pure deflagration of a possibly sub-Chandrasekhar-mass WD
\citep[e.g.,][]{Branch04, Phillips07}.  This scenario explains the low
velocities and luminosities of these events.  Such an event would
retain a significant amount of unburned material in its core, leading
to strong [\ion{O}{1}] emission at late times \citep{Kozma05};
however, this has not been seen in the late-time spectra of SNe~2002cx
or 2005hk \citep[][although this may only be the result of the high
density at late times]{Jha06:02cx, Sahu08}.

In the context of SN~2008ha, the energetics constrain the model
further.  If the progenitor star was a Chandrasekhar-mass WD, the
energy released, $E_{\rm released}$, during the explosion must be
larger than the binding energy, $E_{\rm binding} \approx
10^{51}$~ergs, to unbind the star.  With an expansion velocity of
2000~km~s$^{-1}$ and 1.4 M$_{\sun}$ of ejecta, SN~2008ha would have
$E_{\rm released} \approx E_{\rm kin} = 5 \times 10^{49}$~ergs or
$E_{\rm released} \approx 5 \times 10^{-2} E_{\rm binding}$.  This
scenario requires that the energy created in the explosion be just
above the binding energy of the star.  Since the kinetic energy scales
with the ejecta mass (which scales with the progenitor mass in this
case), a sub-Chandrasekhar explosion does not improve the energy
problem.  This balancing problem means that it is unlikely that
SN~2008ha was a deflagration of a WD that disrupted its progenitor
star.  Furthermore, in \S~\ref{s:energy}, we found $M_{\rm ej}
\approx 0.15$ M$_{\sun}$, significantly below what is expected from a
full disruption of a WD.

Numerical models of WD explosions have found some events which produce
a deflagration burning front that rises from within the star to the
surface, reaching a velocity of \about 5000~km~s$^{-1}$ at the surface
with some material up to \about 10000~km~s$^{-1}$, but does not unbind
the star \citep[e.g.,][]{Calder04, Livne05}.  In this scenario, we
expect relatively small $^{56}$Ni (\about $3 \times 10^{-2}$
M$_{\sun}$) and ejecta masses.

Except for the lack of late-time [\ion{O}{1}] emission, the
higher-luminosity members of the SN~2002cx-like class are consistent
with a full deflagration of a WD; however, SN~2008ha is inconsistent.
On the other hand, the light curves, line velocities, composition,
energetics, and host-galaxy properties of SN~2008ha are all consistent
with a failed deflagration.  If this model is correct for SN~2007J,
the He lines must be the result of interaction with circumstellar
material.

\subsection{.Ia ``Supernova''}

For the ``.Ia'' model, first proposed by \citet{Bildsten07}, a binary
WD system similar to AM CVn with a low accretion rate will eventually
undergo a particularly luminous outburst caused by nuclear burning of
a helium shell with $M \approx 0.05$ M$_{\sun}$.  This flash will
generate Fe-peak radioactive nuclei similar to a SN~Ia, but the
smaller mass of radioactive elements and ejected material produce a
less luminous, rapid light curve, with high-velocity ejecta.

This model makes several predictions.  The $^{56}$Ni mass should be
\about $10^{-2}$ M$_{\sun}$ and the absolute magnitude should be in the
range of $-15$ to $-18$~mag.  Since the ejecta mass should be small
(similar to the mass of the helium shell, 0.05 M$_{\sun}$), the rise
time should be short (2--10~days).  Because there is a large
$M_{^{56}{\rm Ni}} / M_{\rm ej}$ ratio, the ejecta velocity should be
large (\about 15000~km~s$^{-1}$). Given that the progenitors are AM
CVn systems, the host-galaxy population should be dominated by
early-type galaxies.  The rise time and luminosity of SN~2008ha are on
the extreme edges of the ranges found by \citet{Bildsten07}.  However,
the ejecta velocity of SN~2008ha is much lower than that predicted,
and its host galaxy is an irregular galaxy, an unlikely galaxy in
which to find such an event (but since there are old stellar systems
in all galaxies, this alone should not eliminate the model).

If we take the model of \citet{Bildsten07} and modify it so that the
nuclear burning is less efficient, the explosion would produce less
$^{56}$Ni but have a similar amount of ejecta, producing a fainter
event with a lower ejecta velocity, while the rise time would be
approximately the same.  If the kinetic energy is proportional to the
$^{56}$Ni mass, then to match the observed ejecta velocity, we would
expect the $^{56}$Ni to be a factor of \about 50 smaller than that
presented by \citet{Bildsten07}.  This implies $M_{^{56}{\rm Ni}} = 2
\times 10^{-3}$ M$_{\sun}$, similar to what we found from an
examination of the light curve in \S~\ref{s:energy}.

The inefficient version of the .Ia model appears to reproduce the
properties of SN~2008ha.  Since this model involves a He WD, some of
the ejecta (or alternatively, circumstellar material from the mass
transfer) may be unburned He.  This model may explain the He emission
lines present in SN~2007J, but lack of strong [\ion{O}{1}] in the
late-time spectra of SN~2002cx.  However, the ejecta mass and rise
times predicted by this model do not match those observed for most
SN~2002cx-like objects.  Additionally, the progenitor systems of this
model are members of an old stellar population, so we are unlikely to
have discovered these objects exclusively in spiral galaxies. Thus,
this model probably does not apply to all SN~2002cx-like events.


\section{Discussion and Conclusions}\label{s:disc}

SN~2008ha is spectroscopically a member of the SN~2002cx-like class of
SNe, but it is an extreme case.  Its luminosity, kinetic energy, and
total energy released are all perhaps the smallest of any SN observed
(excluding probable LBV outbursts misclassified as SNe;
\citealt{VanDyk03, Maund06, Berger09, Smith09}; and references
therein).  If SN~2008ha was a thermonuclear explosion (either a failed
deflagration or a SN~.Ia), the progenitor star survived the explosion.
If the SN was a core-collapse event, there should be a black hole
remnant.

Until recently, the SN~2002cx-like SNe~Ia have been described as
peculiar SNe~Ia.  The extremely low luminosity and kinetic energy of
SN~2008ha require that we re-examine the accepted ideas regarding the
progenitors of this class.  These objects are hosted in non-elliptical
galaxies, but their host-galaxy distribution is statistically similar
to that of any SN type except for SN~1991bg-like objects.  This
distribution, including two objects found in S0 galaxies, makes the
direct collapse/fallback scenario unlikely for some objects of this
class.  Similarly, the lack of objects in elliptical galaxies makes
the SN~.Ia model unlikely.  Although they do not fit the observations
perfectly, we prefer the electron-capture and deflagration scenarios
for this class.

Our data do not exclude thermonuclear explosions for SN~2008ha or any
SN~2002cx-like object.  Although there are core-collapse models which
are adequate at describing our observations of SN~2008ha, unlike
\citetalias{Valenti09} we do not claim that SN~2008ha (and more
generally, all SN~2002cx-like objects) are definitively core-collapse
events.

The LOSS and amateur SN searches, which have found the vast majority
of the nearby SNe in the last decade and whose depth is typically mag
19, would only be able to detect objects similar to SN~2008ha to a
distance modulus of $\mu \approx 33$~mag, corresponding to 40~Mpc.  In
the past 10 years, there have been 60 SNe~Ia discovered with $\mu \le
33$~mag.  Considering that objects as faint as SN~2008ha would stay
above the detection limit for a very short time in most of these
galaxies, while normal SNe~Ia would be observable for over a year, it
is quite possible that SN~2008ha-like events could represent 2\%--10\%
of SNe~Ia; however, SN~2008ha may be a singular event.  The rough rate
determined from this single object is very similar to the rate
determined by \citet{Bildsten07} for AM CVn outbursts of 2\%--6\% in
late-type galaxies as well as the WD merger rate.

Further observations of SN~2008ha will reveal additional information
about this interesting object.  Continuing imaging will provide a
better comparison to other SN light curves and will constrain models.
A late-time spectrum should reveal detailed information about the
explosion and its byproducts.  Deep early-time X-ray and radio
observations of SN~2002cx-like objects may help distinguish between
models.

Finally, the era of large, deep transient surveys is about to begin.
Searches such as Pan-STARRS, PTF, SkyMapper, and LSST, which should
find $10^{4}$--$10^{6}$ SNe per year, may provide thousands of objects
similar to SN~2008ha over their lifetimes, allowing a statistical
analysis of a large sample of objects.

\begin{acknowledgments} 

R.J.F.\ would like to thank the many people who have contributed in
some way to this work: J.\ Bochanski, A.\ Burgasser, W.\ High, A.\
Rest, C.\ Stubbs, and D.\ Sand, who were willing to observe this
object during their observing runs; C.\ Griffith, M.\ Kislak, J.\
Leja, T.\ Lowe, B.\ Macomber, and P.\ Thrasher, who performed many of
the Nickel observations; C. Griffith, F. Serduke, and X. Wang, who
obtained some spectra with the Lick 3-m reflector; and W.\ Peters, A.\
Vaz, and N.\ Wright, who performed the FAST observations.  We are
grateful to the kind observers at Gemini for obtaining our
observations, to A.\ Soderberg for acquiring the first PANIC points
during time she was observing for M.W.V., and to T.\ Armandroff and
M.\ Kassis for allowing us to observe during engineering time on Keck.
The Gemini director's time observations (program ID GN-2008B-DD-6)
were critical for this study; we particularly thank J.-R.\ Roy and D.\
Crabtree for their approval, and A.\ Stephens and K.\ Volk for their
execution of our program.  Discussions with L.\ Bildsten, B.\ Metzger,
T.\ Piro, and E.\ Quataert improved the section on theoretical models.
Additional discussions with E.\ Berger, D.\ Sand, and A.\ Soderberg
gave insight into the nature of this event.  Software written by G.\
Becker was used in the reduction of our MagE data; his guidance and
discussions regarding his code were greatly appreciated.  T.\ Matheson
rereduced the spectrum of SN~1991bj after nearly two decades of
neglect on an old tape.  J.\ Prieto notified us to a misclassification
in NED for the host of SN~2008ae in an early version of this
manuscript.

Based in part on observations obtained at the Gemini Observatory,
which is operated by the Association of Universities for Research in
Astronomy, Inc., under a cooperative agreement with the US National
Science Foundation on behalf of the Gemini partnership: the NSF
(United States), the Science and Technology Facilities Council (United
Kingdom), the National Research Council (Canada), CONICYT (Chile), the
Australian Research Council (Australia), Minist\'{e}rio da Ci\^{e}ncia
e Tecnologia (Brazil) and Ministerio de Ciencia, Tecnolog\'{i}a e
Innovaci\'{o}n Productiva (Argentina).  Some of the data presented
herein were obtained at the W.~M. Keck Observatory, which is operated
as a scientific partnership among the California Institute of
Technology, the University of California, and the National Aeronautics
and Space Administration (NASA); the observatory was made possible by
the generous financial support of the W.~M. Keck Foundation.  We
acknowledge the use of public data from the {\it Swift} data archive.
We have made use of the SUSPECT SN spectra archive.  We are grateful
to the staffs at the Lick, Keck, Gemini, and Fred L.\ Whipple
Observatories for their dedicated services.  KAIT was constructed and
supported by donations from Sun Microsystems, Inc., the
Hewlett-Packard Company, AutoScope Corporation, Lick Observatory, the
NSF, the University of California, the Sylvia \& Jim Katzman
Foundation, and the TABASGO Foundation.  PAIRITEL is operated by the
Smithsonian Astrophysical Observatory (SAO) and was made possible by a
grant from the Harvard University Milton Fund, the camera loan from
the University of Virginia, and the continued support of the SAO and
UC Berkeley. The PAIRITEL project is supported by NASA/Swift Guest
Investigator Grant NNG06GH50G.  We are especially grateful to J.\
Bloom, D.\ Starr, C.\ Blake, A.\ Szentgyorgyi, and M.\ Skrutskie for
developing and maintaining PAIRITEL and E.\ Falco and the Mt.\ Hopkins
staff (W.\ Peters, R.\ Hutchins, and T.\ Groner) for their continued
assistance with PAIRITEL.  Supernova research at Harvard is supported
by NSF grant AST--0606772.  A.V.F.'s supernova group at U.C. Berkeley
is supported by NSF grant AST--0607485, Gary and Cynthia Bengier,
Richard and Rhoda Goldman, and the TABASGO Foundation.  This research
has made use of the NASA/IPAC Extragalactic Database (NED), which is
operated by the Jet Propulsion Laboratory, California Institute of
Technology, under contract with NASA.  M.M.\ is supported by a
research fellowship from the Miller Institute of Basic Research in
Science at U.C. Berkeley.

{\it Facilities:} 
\facility{FLWO:1.5m(FAST), FLWO:PAIRITEL, Gemini:North(NIRI),
Keck:I(LRIS), Lick:KAIT, Lick:Nickel, Lick:Shane(Kast),
Magellan:Baade(PANIC), Magellan:Clay(MagE), MMT(Blue Channel),
Swift(UVOT, XRT)}

\end{acknowledgments}

\appendix
\section{SN 2007J}\label{s:07j}

SN~2007J was discovered in UGC~1778, which has a recession velocity of
5034~km~s$^{-1}$ \citep{Wegner93} and Galactic coordinates $(l, b) =
(143.0^\circ, -25.7^circ)$ (yielding a distance modulus $\mu =
34.24$~mag), on 2008 Jan. 15.21 at mag 18.2 \citep{Lee07}.  It was
originally classified as a SN~2002cx-like object
\citep{Filippenko07:07J1}, but a later spectrum showed \ion{He}{1}
emission lines, causing \citet{Filippenko07:07J2} to reclassify
SN~2007J as a SN~Ib.

\begin{figure}
\begin{center}
\epsscale{0.4}
\rotatebox{0}{
\plotone{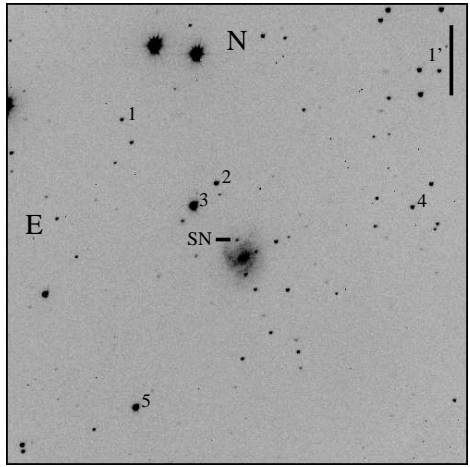}}
\caption{KAIT $R$-band image of SN~2007J and its host galaxy,
UGC~1778. The FOV is 6.7\arcmin $\times$ 6.7\arcmin.  The SN and
comparison stars are marked.  The labels for the comparison stars
correspond to the numbers in
Table~\ref{t:kait_stars_07j}.}\label{f:finder_07j}
\end{center}
\end{figure}

To determine if SN~2007J is truly a member of the SN~2002cx-like
class, we examine its light curve and spectra in more detail.  A
finding chart of SN~2007J, its host galaxy, and comparison stars is
shown in Figure~\ref{f:finder_07j}.

\subsection{Photometry}\label{ss:phot_07j}

\begin{figure}
\begin{center}
\epsscale{0.6}
\rotatebox{90}{
\plotone{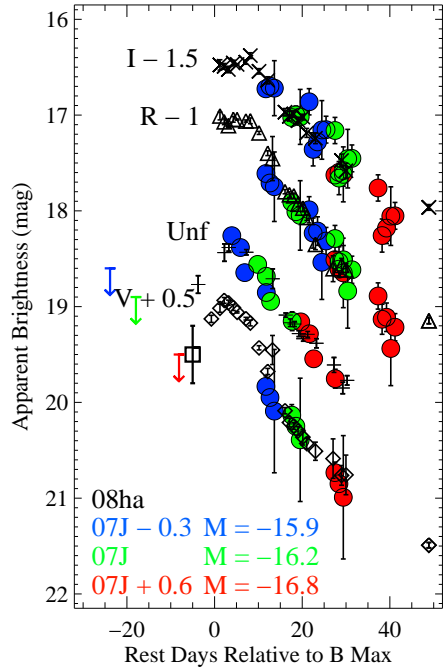}}
\caption{Apparent \vri and unfiltered (labeled ``Unf'' in the figure)
photometry of SN~2007J (green circles) compared to the $V$ (diamonds),
unfiltered (crosses), $R$ (triangles), and $I$ (Xs) of SN~2008ha,
shifted to match the photometry of SN~2007J.  The blue circles
correspond to a shift of $-0.3$~mag and $-6$~days relative to the
green points.  Similarly, the red circles correspond to a shift of
0.6~mag and $+10$~days relative to the green points.  Limits in the
unfiltered band are also plotted.  The blue, green, and red curves
represent the shifts required to match our first unfiltered
observation of SN~2007J to the peak of the unfiltered light curve of
SN~2008ha, to best match the light-curve shape of SN~2008ha, and the
furthest shift in time to reasonably account for our nondetection
limit, respectively.  The blue, green, and red curves correspond to
$M_{\rm unf} = -15.9$, $-16.2$, and $-16.8$~mag at peak, respectively.
The discovery magnitude of SN~2008ha reported by \citet{Puckett08} is
plotted as a square.}\label{f:07j_lc}
\end{center}
\end{figure}

In Figure~\ref{f:07j_lc}, we present our \vri and unfiltered light
curves of SN~2007J matched to SN~2008ha.  Our KAIT photometry is given
in Table~\ref{t:kait_07j} and calibration-star information can be
found in Table~\ref{t:kait_stars_07j}.  In Figure~\ref{f:07j_lc}, we
examine the absolute magnitude of SN~2007J by matching its light
curves to those of SN~2008ha.  We apply three different shifts of
magnitude and in time to roughly match SN~2008ha with modest
constraints.  Despite the poor coverage of SN~2007J, the light curves
appear to be very similar, with both having very fast declines.  If we
assume that our first observation of SN~2007J corresponds to its peak
brightness, we can place an upper limit on the peak absolute magnitude
of SN~2007J.  Similarly, using our nondetection of SN~2007J and
assuming that the light curves of SNe~2007J and 2008ha are similar, we
can place a lower limit on its absolute magnitude at peak.

Assuming a Milky Way extinction of $A_{\rm unf} = 0.21$~mag
\citep{Schlegel98} and negligible host-galaxy extinction, we determine
that the absolute magnitude of SN~2007J falls in the range $-15.8 \ge
M_{\rm unf} \ge -17.2$~mag.  Matching the features in the light curves
of SNe~2007J and 2008ha, which are of low significance, we find
$M_{\rm unf} \approx -16.0$~mag for SN~2007J.

 \begin{deluxetable*}{lccccc}
 \tablewidth{0pc}
 \tablecaption{KAIT Photometry of SN~2007J\label{t:kait_07j}}
 \tablehead{\colhead{JD} & \colhead{$V$ (mag)}& \colhead{$R$ (mag)}& \colhead{$I$ (mag)} & \colhead{Unfiltered (mag)}}

\startdata

2454075.74 & \nodata       &  \nodata      &  \nodata      & $> 19.1$      \\
2454107.69 & \nodata       &  \nodata      &  \nodata      & $> 16.7$      \\
2454115.75 & \nodata       &  \nodata      &  \nodata      & 18.559 (050)  \\
2454117.75 & \nodata       &  \nodata      &  \nodata      & 18.683 (075)  \\
2454118.75 & \nodata       &  \nodata      &  \nodata      & 18.945 (058)  \\
2454123.66 & 19.634  (110) & 18.910  (065) & 18.526  (062) & 19.153 (092)  \\
2454124.61 & 19.749  (092) & 19.010  (058) & 18.498  (094) &  \nodata      \\
2454125.62 & 19.891  (645) & 19.049  (363) & 18.519  (288) &  \nodata      \\
2454133.69 & \nodata       & 19.289  (138) & 18.662  (138) &  \nodata      \\
2454134.62 & \nodata       & 19.532  (158) & 19.159  (174) &  \nodata      \\
2454135.64 & \nodata       & 19.511  (150) & 19.080  (163) &  \nodata      \\
2454136.64 & \nodata       & 19.837  (388) & 18.956  (306) &  \nodata      \\
2454137.62 & \nodata       & 19.615  (143) & 18.952  (140) &  \nodata     

\enddata

\tablecomments{Uncertainties are given in parentheses.}

 \end{deluxetable*}

 \begin{deluxetable*}{lccccccc}
 \tablewidth{0pc}
 \tablecaption{Comparison Stars for SN~2007J\label{t:kait_stars_07j}}
 \tablehead{\colhead{Star} & \colhead{$\alpha$(J2000)} &\colhead{$\delta$(J2000)} &\colhead{$B$ (mag)} & \colhead{$V$ (mag)} & \colhead{$R$ (mag)} & \colhead{$I$ (mag)} & \colhead{$N_{\rm calib}$}}
\startdata
 SN & 02:18:51.70 & +33:43:43.3 & & & & & \\
  1 & 02:18:59.65 & +33:45:28.1 &   18.681(008) &   17.830(017) &   17.379(004) &   16.939(004) &  3\\
  2 & 02:18:53.14 & +33:44:32.7 &   18.481(016) &   17.105(013) &   16.173(005) &   15.408(007) &  3\\
  3 & 02:18:54.71 & +33:44:13.4 &   14.365(005) &   13.785(011) &   13.434(001) &   13.092(009) &  3\\
  4 & 02:18:39.59 & +33:44:11.6 &   18.609(003) &   17.552(015) &   16.928(006) &   16.455(008) &  3\\
  5 & 02:18:58.77 & +33:41:19.5 &   16.005(005) &   15.119(011) &   14.618(001) &   14.134(006) &  3\\
\enddata
\tablecomments{Uncertainties are given in parentheses.}

 \end{deluxetable*}

We have assumed no color difference between SNe~2007J and 2008ha.
When shifting to determine the luminosity limits, the colors are
slightly different from SN~2008ha, while the best-fit shift has
essentially the same colors as SN~2008ha.  Although SN~2007J appears
to have a lower luminosity than normal SNe~Ia, and may have a
luminosity below most SN~2002cx-like SNe and normal SNe~Ib/c (although
the limits are consistent with SN~2007J having a luminosity similar to
those classes), it is still much more luminous than SN~2008ha which
peaked at $M_{\rm unf} = -14.1$~mag.  As we discuss below, SN~2007J is
spectroscopically similar to SN~2002cx, but develops \ion{He}{1} lines
at late times.  Our light curves of SN~2007J do not cover this epoch.

\subsection{Spectroscopy}\label{ss:spec_07j}

\begin{figure}
\begin{center}
\epsscale{0.4}
\rotatebox{90}{
\plotone{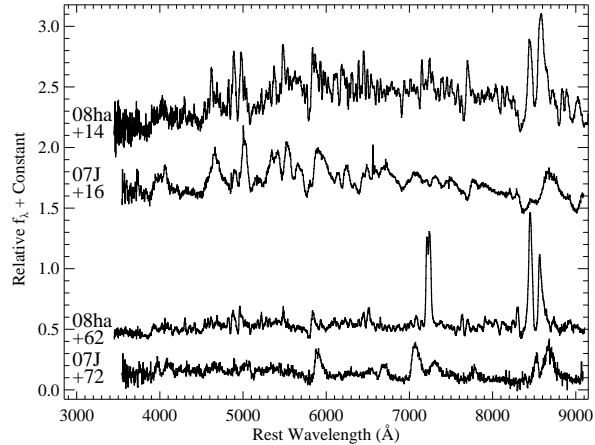}}
\caption{Optical spectra of SNe~2007J and 2008ha.  The spectra are
denoted by their phase relative to maximum brightness in the
unfiltered (for SN~2007J) and $B$ (for SN~2008ha) bands.  We have
blueshifted the spectra of SN~2008ha (after removing its recession
velocity) by a velocity of 3000~km~s$^{-1}$.  For clarity, we have
interpolated over nebular emission lines from the host
galaxy.}\label{f:07j_spec}
\end{center}
\end{figure}

From our analysis of its light curve above, we see that our SN~2007J
spectra obtained on 2007 Jan. 21.4 and 2007 Mar.  18.3 correspond
roughly to phases of $+16$ and $+72$~days, respectively.  In
Figure~\ref{f:07j_spec}, we present our spectra of SN~2007J compared
to spectra of SN~2008ha at similar phases.

Our first spectrum of SN~2007J is fairly similar to that of SN~2008ha
at a similar epoch (after blueshifting the spectrum of SN~2008ha by
3000~km~s$^{-1}$), displaying most of the same line features.  Even
ignoring the differences in line velocity and width, there are
additional differences between these spectra.  Our second spectrum of
SN~2007J is less similar to that of SN~2008ha.  Both spectra contain
many of the same features, but SN~2007J now shows prominent
\ion{He}{1} emission lines while SN~2008ha has strong \ion{Ca}{2} and
[\ion{Ca}{2}] emission lines.  Although there are several differences
between the objects, SN~2007J is relatively similar to SN~2008ha and
is a potential member of the SN~2002cx-like class.

The \ion{He}{1} emission present in the spectrum of SN~2007J caused
\citet{Filippenko07:07J2} to argue that SN~2007J was a core-collapse
event.  However, \ion{He}{1} emission by itself simply indicates He in
the SN environment, not necessarily that the progenitor underwent core
collapse.  The \ion{He}{1} line with the least contamination from
other lines in our SN~2007J spectrum is the $\lambda 7065$ line; with
a FWHM of 3900~km~s$^{-1}$, it is wider than other lines in the
spectra (typical FWHM $\approx 1500$~km~s$^{-1}$).  The \ion{He}{1}
line profiles are similar to those of the peculiar SN~Ib~2006jc
\citep{Foley07, Pastorello07}, perhaps indicating that the \ion{He}{1}
emission is the result of circumstellar interaction rather than
forming in the SN photosphere.  \ion{He}{1} emission lines have not
been seen in the spectra of any other SN~2002cx-like SNe, including
very late-time spectra \citep{Jha06:02cx, Sahu08}, casting the
SN~2002cx-like classification for SN~2007J in doubt.

\bibliographystyle{apj}
\bibliography{astro_refs}


\end{document}